\documentclass[english,useAMS, usenatbib]{mn2e}
\usepackage[T1]{fontenc}
\usepackage[latin9]{inputenc}
\usepackage{units}
\usepackage{rotating}
\usepackage{url}
\usepackage{amsmath}
\usepackage{commath}
\usepackage{amssymb}
\usepackage{graphicx}
\usepackage{subfig}
\usepackage{pdfpages}
\usepackage{esint}
\usepackage[authoryear]{natbib}
\usepackage{listings}
\usepackage{textcomp}
\usepackage[export]{adjustbox}
\usepackage{booktabs}

\makeatletter

\def\gtsima{$\; \buildrel > \over \sim \;$}
\def\ltsima{$\; \buildrel < \over \sim \;$}
\def\prosima{$\; \buildrel \propto \over \sim \;$}
\def\gsim{\lower.5ex\hbox{\gtsima}}
\def\lsim{\lower.5ex\hbox{\ltsima}}
\def\simgt{\lower.5ex\hbox{\gtsima}}
\def\simlt{\lower.5ex\hbox{\ltsima}}
\def\simpr{\lower.5ex\hbox{\prosima}}
\def\la{\lsim}
\def\ga{\gsim}

\newcommand{\be}{\begin{eqnarray}}
\newcommand{\ee}{\end{eqnarray}}
\def\lsim{\,\lower2truept\hbox{${< \atop\hbox{\raise4truept\hbox{$\sim$}}}$}\,}
\def\gsim{\,\lower2truept\hbox{${> \atop\hbox{\raise4truept\hbox{$\sim$}}}$}\,}

\providecommand{\tabularnewline}{\\}

\title[Dust in high redshift galaxies]{The assembly of dusty galaxies at $z \geq 4$: statistical properties}
\author[Graziani et al.]{L. Graziani$^{1,2,3}$\thanks{E-mail:
luca.graziani@roma1.infn.it}, R. Schneider$^{1,2,4}$, M. Ginolfi$^{5}$, L. K. Hunt$^{3}$, U. Maio$^{6}$, 
\newauthor M. Glatzle$^{7,8}$, B. Ciardi$^{8}$\\
$^{1}$Dipartimento di Fisica, Sapienza, Universit$\grave{a}$ di Roma, Piazzale Aldo Moro 5, 00185, Roma, Italy\\
$^{2}$INFN, Sezione di Roma I, P.le Aldo Moro 2, 00185 Roma, Italy\\
$^{3}$INAF/Osservatorio Astrofisico di Arcetri, Largo E. Femi 5, 50125 Firenze, Italy \\
$^{4}$INAF/Osservatorio Astronomico di Roma, Via di Frascati 33, 00078 Monte Porzio Catone, Italy\\
$^{5}$Observatoire de Gen\`eve, Universit\'e de Gen\`eve, 51 Ch. des Maillettes, 1290 Versoix, Switzerland \\
$^{6}$Leibniz-Institut f\"ur Astrophysik, An der Sternwarte 16, D-14482 Potsdam, Germany \\
$^{7}$Physik-Department, Technische Universit\"at M\"unchen, James-Franck-Str. 1,85748 Garching, Germany \\
$^{8}$Max-Planck-Institut f\"ur Astrophysik, Karl-Schwarzschild-Stra{\ss}e 1, D-85748 Garching b. M\"unchen, Germany
}

\begin{document}

\date{July 2019}

\pagerange{\pageref{firstpage}--\pageref{lastpage}} \pubyear{2016}

\maketitle

\label{firstpage}

\begin{abstract}

The recent discovery of high redshift dusty galaxies implies a rapid dust enrichment of their interstellar medium (ISM). To interpret these observations, we run a cosmological simulation in a 30$h^{-1}$~cMpc/size volume down to $z \approx 4$. We use the hydrodynamical code \texttt{dustyGadget}, which accounts for the production of dust by stellar populations and its evolution in the ISM.  We find that the cosmic dust density parameter ($\Omega_{\rm d}$) is mainly driven by stellar dust at $z \gtrsim 10$, so that mass- and metallicity-dependent yields are required to assess the dust content in the first galaxies. At $z \lesssim 9$ the growth of grains in the ISM of evolved systems (Log$(M_{\star}/M_{\odot})>8.5$) significantly increases their dust mass, in agreement with observations in the redshift range $4 \lesssim z < 8$. Our simulation shows that the variety of high redshift galaxies observed with ALMA can naturally be accounted for by modeling the grain-growth timescale as a function of the physical conditions in the gas cold  phase. In addition, the trends of dust-to-metal (DTM) and dust-to-gas (${\cal D}$) ratios are compatible with the available data. A qualitative investigation of the inhomogeneous dust distribution in a representative massive halo at $z \approx 4$ shows that dust is found from the central galaxy up to the closest satellites along polluted filaments with $\rm Log({\cal D}) \leq -2.4$, but sharply declines at distances $d \gtrsim 30$~kpc along many lines of sight, where $\rm Log({\cal D}) \lesssim -4.0$.    

\end{abstract}


\section{Introduction}

Recent observations performed with the {\it Atacama Large Millimeter Array} (ALMA)\footnote{http://www.almaobservatory.org} have confirmed the dusty nature of ``normal'' star forming galaxies\footnote{In this paper, normal galaxies are identified as non-starburst objects with star formation rates of a few tens of solar masses per year, representing the dominant class of galaxies at early cosmic times.} at early epochs ($z \geq 4$). Dust continuum detections \citep{2015Natur.519..327W, 2017MNRAS.466..138K, 2017ApJ...837L..21L}, upper limits \citep{2015A&A...574A..19S, 2015MNRAS.452...54M, 2016ApJ...833...71A} and line emissions \citep{2017ApJ...836L...2B, 2016Sci...352.1559I,2017arXiv170804936O} are now available for a limited sample of these chemically evolved systems (see \citealt{2014PhR...541...45C} for a recent review); however, their number will certainly increase with future ALMA programs, with the advent of the {\it James Webb Space Telescope} (JWST)\footnote{http://www.jwst.nasa.gov/} and with the {\it Extremely Large Telescope} (ELT)\footnote{http://www.eso.org/public/teles-instr/elt/}.
 
To understand the evolution of these galaxies in the epoch of hydrogen reionization \citep{2018MNRAS.477..552B} and to interpret their observables from ab-initio physical properties \citep{2016MNRAS.462.3130M, 2017MNRAS.470.3006C}, theoretical models of galaxy formation accounting for radiative and chemical feedback have been recently developed \citep{2012ApJ...745...50W, 2015MNRAS.449.3137G,xu2016,2017MNRAS.469.1101G, 2017MNRAS.471.4128P, 2018MNRAS.480.4842C, 2019MNRAS.482..321G, 2019MNRAS.tmp.1637K}. They are of strategic importance to describe the multi-phase, metal enriched ISM of these early galaxies \citep{2003ApJ...587..278W, 2013ARA&A..51..105C} and their circumgalactic/intergalactic medium (CGM/IGM) \citep{2018MNRAS.480.2628F}. On a cosmological scale, these models can shed light on the impact of cosmic dust on the high-redshift luminosity function \citep{2016ApJ...833..254S, 2017MNRAS.471.4155K, 2017arXiv170406004O}, on the early stages of cosmic reionization \citep{2018MNRAS.476.1174E} and on the colors of galaxy populations \citep{2013MNRAS.432.3520D}.

Over the last years, improvements have been made in the chemical network of semi-analytic \citep{2017MNRAS.465..926D,  2017MNRAS.471.3152P,2019arXiv190402196V}, semi-numerical \citep{2015MNRAS.451L..70M, 2016MNRAS.455..659W, Khakhaleva2016,2016ApJ...831..147Z,2017arXiv170505858N,2018MNRAS.473.4538G}, and numerical models of galaxy formation \citep{2015MNRAS.449.1625B,2015ApJ...799..166B,2017MNRAS.468.1505M, 2017MNRAS.466..105A,2018MNRAS.479.2588G, 2018MNRAS.478.1694M} but despite these advancements, the introduction of a comprehensive treatment of cosmic dust in cosmological simulations remains extremely challenging. 

The origin and composition of dust grains is highly uncertain and models of dust nucleation in supernova (SN) ejecta \citep{2004MNRAS.351.1379S,2007MNRAS.378..973B,2015MNRAS.454.4250M, 2015A&A...575A..95S,2016A&A...587A.157B, 2016arXiv161209013S,2019MNRAS.484.2587M} and in the atmosphere of Asymptotic Giant Branch (AGB) stars \citep{ferrarotti2006, zhukovska2008, 2012MNRAS.420.1442V, 2012MNRAS.424.2345V, 2013MNRAS.434.2390N, nanni2014, 2013MNRAS.433..313D, 2017MNRAS.467.4431D, ventura2018, dellagli2019} are required. So far, stellar dust yields 
adopted in cosmological simulations remain highly unconstrained or model dependent, especially for massive stars with low initial metallicity. In the first galaxies, 
accurate yields are required to account for the release of metals by the first stars (Pop\ III) \citep{2003ApJ...598..785N,2004MNRAS.351.1379S,2014ApJ...794..100M,2015MNRAS.454.4250M,2018ApJ...857..111T,2019MNRAS.482.3933C}, and to understand the transition from Pop\ III to successive generations (Pop\ II) \citep{2010MNRAS.407.1003M,schneider12b,schneider12a, 2014MNRAS.445.3039D, 2015MNRAS.446.2659C}. Even the composition of dust in our Galaxy and in its Local Group companions remains a subject of debate, because of uncertainties in interpreting depletion of atomic metals along local lines of sight \citep{1994ApJ...424..748C,2004ApJ...605..272S}, the variety of grain chemical compositions, and the difficulties in modeling the observed extinction curves, often contaminated by molecules \citep{2003ARA&A..41..241D, 2003ApJ...594..279G,2015ApJ...815...14C,2019MNRAS.487.1844M}.  

There is observational evidence that dust grains undergo modifications depending on the ISM phase where they
reside. Grains can sublimate in extremely hot environments, or can be destroyed by a number of processes such as 
shattering in grain-grain collisions and  thermal sputtering\footnote{  The interested reader is referred to  \citet{2011piim.book.....D} and references therein.}. Shocked gas fronts, propagating in the ISM as a result of supernova explosions, are candidate environments in which the above processes act efficiently \citep{1994ApJ...433..797J,1995Ap&SS.233..111D, 1997A&A...322..296C}. Where amorphous dust grains can survive, they significantly evolve by changing their physical properties: mass, size \citep{2012MNRAS.422.1263H,2017ApJ...841...72R}, chemical composition \citep{2010MNRAS.408..535C} and charge \citep{2001ApJS..134..263W,2004ASPC..309..453W}. These grains can even morph into crystals, if an intense UV flux is present \citep{2013A&A...558A..62J}. Depending on the problem at hand and its physical scale and considering the computational cost, numerical implementations generally account for only a subset of the above processes.

Dust production by SNe and AGB stars, as well as processes of grain evolution in the galactic ISM, were implemented in SPH and grid-based schemes \citep{2015MNRAS.449.1625B, 2015ApJ...799..166B, 2017MNRAS.468.1505M, 2017MNRAS.466..105A}, while other codes focused on dust feedback in momentum driven winds \citep{2014MNRAS.444.3879B,2014MNRAS.445..581H, 2016MNRAS.456.4174H}, or on computing the radiation extinction by radiative transfer through a dusty ISM \citep{2000ApJ...545...86W, 2011MNRAS.412.1059Z, 2013ApJ...776...35K, 2014MNRAS.440..134A, 2016PASJ...68...94H}.

In \citet{2015MNRAS.451L..70M}, we introduced a novel semi-numerical model of dusty galaxies by coupling the results of 
a semi-analytic code \citep{2014MNRAS.445.3039D} with a SPH simulation \citep{2010MNRAS.407.1003M}, and we first interpreted the dust mass of normal, high-redshift ($z>6$) galaxies as a result of production by stars and  efficient grain growth in the dense phases of the ISM. The successive coupling with a semi-analytic treatment of radiative feedback allowed us to explain the evolution of galaxy colours \citep{2016MNRAS.462.3130M} and to show that current high redshift observations already provide important constraints on the nature of dust and its complex evolution in the various phases of the ISM. 

Here we go a step forward by introducing a numerical implementation of our dust model in the cosmological code \texttt{Gadget} \citep{2005MNRAS.364.1105S} and successive extensions \citep{2007MNRAS.382..945T,2007MNRAS.382.1050T,2009A&A...503...25M}. The new code, named \texttt{dustyGadget}, implements a consistent evolution of grains in different phases of the ISM and follows the spreading of dust and atomic metals by galactic winds throughout the scales of circum- and intergalactic medium (CGM/IGM). 

The paper is organized as follows. In Section \ref{sec:dustGadget} we describe the numerical implementation of our code, while the set-up of the galaxy formation simulation is provided in Section \ref{sec:galForm}. Section \ref{sec:ThModels} introduces the theoretical models used to benchmark the findings of \ref{sec:galForm}. Finally, the simulation results are in Section \ref{sec:results}: the redshift evolution of {the dust content} is described in  \ref{sec:dustZ} and the statistics of our galaxy sample are discussed in \ref{sec:galZoo} and carefully compared to current observations at $z \ge 4$. A qualitative analysis of the spatial distribution of dust in a massive dusty halo at $z\approx4$ is the subject of Section \ref{sec:dustyMMHalo}. The results of our investigation are summarized and discussed in Section \ref{sec:dustConclusions}.

\section{\texttt{dustyGadget}}
\label{sec:dustGadget}

Here we describe the main features of \texttt{dustyGadget} and their numerical implementation. Section \ref{sec:chemistry} and Appendix \ref{sec:AppAISMModel} summarize the chemical network and the ISM model inherited from previous implementations, while Sections  \ref{sec:dustproduction} and \ref{sec:dustevolution} focus on dust production by stars and the evolution of grains in the multiphase ISM.

\subsection{Chemical network: atomic metals and molecules}
\label{sec:chemistry}

\texttt{dustyGadget} derives its gas chemical evolution model from the original implementation of \citet{2007MNRAS.382.1050T}. The model relaxes the so-called {\it Instantaneous Recycling Approximation} (IRA) and follows the metal release from stars of different masses, metallicity and lifetimes \citep{1993ApJ...416...26P}. Different models of the adopted  metal yields, as well as alternative Initial Mass Functions (IMF) or functional forms of the adopted stellar lifetime can also be easily implemented in the code and their impact on the results explored \citep{2005A&A...430..491R,2010A&A...522A..32R, 2016MNRAS.455.4183V}. Mass and metallicity-dependent yields are implemented for Pop\ II/I stars: for low and intermediate mass stars we adopt \citet{1997A&AS..123..305V}, while results from \citet{1995ApJS..101..181W} are included to describe core-collapse SNe. Finally, for type-Ia supernovae (SNIa) we use \citet{2003NuPhA.718..139T}. Stars with masses $\geq 40~M_\odot$ are assumed to collapse into black holes and do not contribute to metal enrichment. Pop\ III stars with masses in the range $[140, 260]~M_{\odot}$ are expected to explode as pair-instability SNe (PISN) and their mass-dependent yields are taken from \citet{2002ApJ...567..532H}. Outside this range, they are assumed to directly collapse into black holes. The chemical network present in \texttt{dustyGadget} also includes the evolution of both atomic and ionized hydrogen, helium and deuterium, as described in \citet{2003ApJ...592..645Y},\citet{2007MNRAS.382.1050T}, \citet{2010MNRAS.407.1003M}. Reactions leading to the creation or destruction of primordial molecules are also included following \citet{2007MNRAS.379..963M} and the gas cooling function consistently  reflects the chemical network by accounting for molecular, atomic, and  fine structure metal transitions of O, $\textrm{C}^{+}$, $\textrm{Si}^{+}$, $\textrm{Fe}^{+}$ at $T<10^{4}\textrm{K}$ \citep{1993ApJS...88..253S, 2003ApJ...592..645Y,2007MNRAS.379..963M,2009MNRAS.393...99W}.

\subsection{Dust production by stars}
\label{sec:dustproduction}

Dust production by stars is implemented to ensure consistency with gas phase metal enrichment: mass and metallicity-dependent dust yields, consistent with the ones described in \ref{sec:chemistry}, are computed for  different stellar populations. Hence, in this first implementation of  \texttt{dustyGadget}, dust yields for Pop III PISN are taken from \citet{2004MNRAS.351.1379S}, which in turn have been computed from the grid of PISN models by \citet{2002ApJ...567..532H}. For Pop II/I core-collapse SNe we use the yields from \citet{2007MNRAS.378..973B} (based on the supernova models by  \citealt{1995ApJS..101..181W}), while for AGB stars we adopt the yields from \citet{2006A&A...447..553F} and \citet{2008A&A...479..453Z} (derived from the models of  \citealt{1997A&AS..123..305V}). However, alternative sets of consistent metal and dust yields could be adopted in future works to explore the impact of more recent calculations of core-collapse SNe \citep{2019MNRAS.484.2587M} and AGB dust yields \citep{2012MNRAS.424.2345V,2012MNRAS.420.1442V,2013MNRAS.433..313D, 2017MNRAS.467.4431D, ventura2018, dellagli2019}.
It would be possible to also explore the impact of different mass ranges and slopes of the Pop III IMF that are still poorly constrained by numerical simulations \citep{hirano2014, hirano2015} and stellar archaeology studies \citep{2017MNRAS.465..926D}.

An important aspect that has to be considered when evaluating supernova dust yields is the effect of the reverse shock (RS) on the mass of dust nucleated in the supernova ejecta \citep{2007ApJ...666..955N, 2007MNRAS.378..973B,2010ApJ...715.1575S, 2012ApJ...748...12S, 2015MNRAS.454.4250M, 2016A&A...587A.157B, 2016A&A...590A..65M}. The concerted impact of thermal and non thermal sputtering by collisions with energetic gas particles in shocked regions can lead to the erosion of dust grains on time scales of $\sim 10^5$~yrs. 
Depending on the density of the circum-stellar medium (typically ranging in $\rho_{\rm ISM} = [10^{-25} - 10^{-23}] ~\rm gr/cm^3$) \citet{2007MNRAS.378..973B} find that only a fraction of the dust mass (from 2 to 20\%) survives the passage of the RS. This is a significant reduction that can not be neglected when estimating the contribution of SNe to interstellar dust enrichment. In a more recent study, \citet{2016A&A...587A.157B} compare the dust masses inferred from observations of four well studied SN remnants in the Milky Way and Magellanic Clouds (SN 1987A, CasA, the Crab nebula, and N49) by adopting  theoretical models which self-consistently follow the dynamics of the grains and account for the effects of the forward and reverse shocks. For all the simulated models, they predict the time evolution of the dust mass in the shocked and un-shocked regions of the ejecta and find good agreement with the values estimated from observations (see their Figure 4). However, since the oldest SN has an estimated age of 4800~yrs and the largest dust mass destruction is predicted to occur between $10^3$ and $10^5$~yrs after the explosions, current observations can only provide an upper limit on the average/effective dust yield of about $(1.55 \pm 1.48) \times 10^{-2} M_\odot$; this is in good agreement with the estimates of  \citet{2007MNRAS.378..973B} for a moderate destruction efficiency (or, equivalently, a circumstellar medium density of $\rho_{\rm ISM} = 10^{-24}~\rm gr/cm^3$). As the RS acts on spatial and temporal scales smaller than the cosmological ones, its impact is accounted for by an {\it effective} dust yield, as described in the above models. 

The tables of stellar dust yields adopted in \texttt{dustyGadget} are in agreement with findings by previous studies \citep{2009MNRAS.397.1661V, Valiante2011, Valiante2014, 2014MNRAS.445.3039D} allowing us to safely compare numerical results across different modelling strategies and over samples of galaxies at high \citep{2015MNRAS.451L..70M, 2016MNRAS.462.3130M} and low \citep{2018MNRAS.473.4538G} redshifts. 
Although yields describing the mass and size distribution of individual dust species are available, in the current implementation four classes are accounted for: Carbon, Silicates (MgSiO$_3$, Mg$_2$SiO$_4$, and SiO$_2$), Alumina (Al$_2$O$_3$) and Iron (Fe) dust grains. Our chemical evolution model is, however sufficiently flexible to include other grain types and to explore combinations of stellar yields and different assumptions on the shapes of the stellar IMF. This is particularly important when dealing with the first phases of dust enrichment operated by Pop\ III SN explosions as the shape of the IMF and the properties of these events are still very uncertain \citep{2017MNRAS.465..926D}. 

As a final remark, we point out that \texttt{dustyGadget} does not explicitly follow the evolution of the grain size distribution once the grains enter the ISM. An explicit computation has been shown to be very expensive \citep{2013MNRAS.432..637A, mckinnon2018}, while a simplified treatment based on a two-size approximation \citep{2015MNRAS.447.2937H} can easily reproduce the main features when implemented in hydrodynamical simulations \citep{2017MNRAS.466..105A}. We plan to include the above approximation in a future work. 

\subsection{Dust evolution in the ISM}
\label{sec:dustevolution}

Once the grains produced by stars are released into the ISM, depending on the environment, they experience a number of physical processes altering their chemical composition, size, charge and temperature. While dust-to-light interactions (e.g. photo-heating, grain charging) do not change the mass of dust unless the grain temperature reaches the sublimation threshold (typically $T_{d,s}  \gtrsim 10^3$~K), other mechanical (i.e. sputtering) and chemical  feedback (i.e. grain growth) can alter both the total mass and grain size distribution \citep{2011piim.book.....D, 2003ARA&A..41..241D}\footnote{In addition, some other mass conserving processes, such as grain coagulation and shattering (i.e. fragmentation by grain-grain collisions), can have profound implications on the grain size distribution.}. A complete dust model should provide a self-consistent evolution of both total dust mass and grain properties (size and temperature at least, see the thorough review by \citealt{2013lcdu.confE..27H}). In this first study we do not follow the evolution of grain sizes but only consider the physical processes which directly alter the dust mass: astration, grain growth, destruction by interstellar shocks and grain sputtering in the hot ISM phase. 

Our implementation relies on the widely adopted multiphase ISM model introduced by \citet{2003MNRAS.339..289S} in \texttt{Gadget2}. Appendix \ref{sec:AppAISMModel}  summarizes its features, while the interested reader can find more details in the original paper.

\texttt{dustyGadget} accounts for dust evolution by implementing the diffuse and condensed phases of our semi-analytic models  \citep{2014MNRAS.445.3039D,2015MNRAS.451L..70M, 2016MNRAS.462.3130M} on top of the hydrodynamical ISM scheme. In each star forming SPH particle a two phase ISM is assumed to structure in hot and cold phases, which are equivalent to the diffuse and condensed phases mentioned above. In this way, at each time step $dt$, equations (7-8) of \citet{2016MNRAS.462.3130M} simply apply, by adapting the nomenclature "diff" $\rightarrow _h$  and "MC" $\rightarrow _c$: 

\begin{equation}
\begin{array}{lcl} \dot{M}_{\rm d,c} & = & \dot{M}^\mathcal{D}_{\rm c}(t) - {\rm SFR}(t)\,\mathcal{D}_{\rm c} + \frac{M_{\rm d,c}(t)}{\tau_{\rm gg}} \\ \dot{M}_{\rm d,h} & = & -\dot{M}^{\mathcal{D}}_{\rm c}(t) +\dot{Y}_{\rm d}(t) - \frac{M_{\rm d,h}(t)}{\tau_{\rm d}}, 
\end{array}
\end{equation}
\noindent
where ${\rm SFR}(t)$ is the time-dependent star formation rate  and $\mathcal{D} = M_{\rm d}/M_{\rm g}$ is the particle mass fraction in dust. $\tau_{\rm d}$ and $\tau_{\rm gg}$ are the dust destruction and accretion timescales, respectively, and $\dot{M}^\mathcal{D}_{\rm c}$ describes the dust mass exchange between the hot phase and the cold phase. Finally, $\dot{Y}_{\rm d}$ is the dust yield produced by the stellar sources and it depends on the SFR, the IMF and the adopted type of metal and dust yields of the specific simulation run (see discussion in Section~\ref{sec:dustproduction}). First, in a numerical scheme featuring an explicit multiphase model, the exchange term $\dot{M}^\mathcal{D}_{\rm c}$ can be directly derived from the mass present in each phase as established by equation~\ref{eq:massexchange}. Second, the gas-to-dust ratio $\mathcal{D}_c$ is computed through the cold gas fraction ($x_c$). With the above nomenclature the evolution of the dust mass in each SPH gas particle becomes: 

\begin{equation}
\dot{M}_{\rm d}  =   - {\rm SFR}(t)\,\mathcal{D}_{\rm c} + \frac{x_{\rm c} M_{\rm d}}{\tau_{\rm gg}} -(1-x_{\rm c}) M_{\rm d}\left(\frac{1}{\tau_{\rm d}}+\frac{3}{\tau_{\rm sp}}\right)+\dot{Y}_{\rm d}(t). 
\label{eq:dustEvolution}
\end{equation}

The interpretation of this equation is straightforward: during a time step $dt$, each star forming SPH particle evolves by losing dust mass through astration (${\rm SFR}\,\mathcal{D}_{\rm c}$), grain destruction ($(1-x_{\rm c}) M_{\rm d}/\tau_{\rm d}$) and sputtering ($(1-x_{\rm c}) M_{\rm d}/\tau_{\rm sp}$); at the same time it gains mass by stellar evolution ($\dot{Y}_{\rm d}$) and grain growth ($ x_{\rm c} M_{\rm d}/\tau_{\rm gg}$). Note that in this final expression we explicitly added a sputtering term (with typical time scale $\tau_{\rm sp}$) to the destruction processes, and it applies to all dust-contaminated particles present in the hot gas phase (also see details of eq. 7). 

Here we recap how the dust mass of a single star forming gas particle is evolved in each timestep $dt$. First, the cold fraction from the previous iteration is used to grow the dust mass in the metal enriched cold gas on the time scale $\tau_{\rm gg}$. Second, as the gas cools down and forms stars, the astration term traps dust into newly formed stellar particles. Finally, SN enrichment injects new grains in the hot phase (the dust mass is accounted for by stellar yields described in \ref{sec:dustproduction}) while shocks destroy on in the time scale $\tau_{\rm d}$. Note that in the hot phase the sputtering term erodes grains on time scale $\tau_{\rm sp}$. 

By iterating this process consistently with the chemical scheme, the mass of the various dust species evolves in time in SPH particles.

\subsubsection{Grain destruction by shocks}
\label{sec:dustDestruction}
Dust grains can be destroyed by sputtering or shattering in hot ISM regions running over supernova shocks. The dust destruction timescale $\tau_{\rm d}$ is modeled as:
\begin{equation}
\tau_{\rm d} = \frac{M_{\rm h}}{R_{\rm SN}^\prime \epsilon_{\rm d} M_{\rm s}(v_{\rm s})}, 
\end{equation}
\noindent
\citep{Valiante2011,2014MNRAS.445.3039D}, where for core-collapse supernovae, 
\begin{equation}
M_{\rm s}(v_{\rm s}) = \frac{\left\langle E_{\rm 51} \right\rangle}{\left\langle v^2_{\rm SN} \right\rangle} = 6800 M_{\odot} \langle E_{51} \rangle /(v_{\rm SN}/(100 {\rm km s^{-1}})^2
\end{equation}
\noindent
is the mass shocked up to a velocity of at least $v_{\rm SN}$ by a SN in the Sedov-Taylor phase. By adopting ${\left\langle v_{\rm SN} \right\rangle \sim 200}$~km/s and $\langle E_{51} \rangle \approx 1.2$ as the average SN energy in units of $10^{51}$~erg, we obtain a typical value of $M_{\rm s}(v_{\rm s}) \approx 2040 M_{\odot}$.

$R_{\rm SN}^\prime$ is the {\it effective} SN rate, since not all SNe are equally efficient at destroying dust \citep{McKee1989}; it is defined by scaling the total supernova rate ($R_{\rm SN}$) of the gas particle with a suitable factor $f_{\rm SN}$: $R_{\rm SN}^\prime = f_{\rm SN} R_{\rm SN} \approx 0.15 R_{\rm SN}$. Finally, the value assumed for the dust destruction efficiency is $\epsilon_{\rm d} = 0.48$ \citep{Nozawa2006}.

For PISN we adopt, in the same equations: $\epsilon_{\rm d} = 0.60$, $f_{\rm PISN} = 1$,  $\langle E_{51} \rangle \approx 27$, and then $M_{\rm s}(v_{\rm s}) \approx 4.59 \times 10^4 M_{\odot}$. The resulting grain destruction time scales are then:
\begin{equation}
\begin{array}{lcl} \tau_{\rm d, SN} & = & 6.8 \times 10^{3}\left(\frac{M_{\rm h}}{10^6 M_{\odot}} \right)\left( \frac{1}{R_{\rm SN}} \right) \\ \tau_{\rm d, PISN} & = & 36.3 \left(\frac{M_{\rm h}}{10^6 M_{\odot}} \right)\left( \frac{1}{R_{\rm PISN}} \right). 
\end{array}
\end{equation}
\noindent
The numerical implementation of these formulas is straightforward in SPH: each time a gas particle is evaluated for stellar evolution, the types of exploding supernova are accounted for, their rates derived, and the mass in hot phase computed after explosion\footnote{It should be noted that in cosmological simulations the propagating environment of SN shocks is likely represented by the same SPH particle, while in zoom-in simulations a criterion to define the affected environment must be adopted, as explained in \citet{2017MNRAS.466..105A}.}.

\subsubsection{Grain growth}

In the cold ISM phase dust can grow in mass by sticking atomic metals onto grain surfaces. While the atomic process is not fully understood from its chemical principles and strictly depends on both environment properties and grain chemical composition and sizes \citep{2018MNRAS.476.1371C}, a commonly adopted parametrization  of the grain growth time scale $\tau_{\rm gg}(n_c,T_c, Z_c)$ is:

\begin{eqnarray}
\tau_{\rm gg} =	2.0 \, {\rm Myr} \times \left(\frac{n_{\rm c}}{1000 \, {\rm cm}^{-3}}\right)^{-1} 
					\left(\frac{T_{\rm c}}{50 \, {\rm K}}\right)^{-1/2}
					\left(\frac{Z_c}{Z_\odot}\right)^{-1}\\ \nonumber
					=\tau_{\rm gg,0}(n_c,T_c) \left(\frac{Z_c}{Z_\odot}\right)^{-1},					 
 \label{eq:tauaccr}
 \end{eqnarray}
 \noindent
where grains are assumed spherical with a typical size of $\approx 0.1 \mu {\rm m}$ \citep{Hirashita2014} and $n_{\rm c}$, $T_{\rm c}$ are the number density and temperature of the cold gas phase. $Z_{\rm c}$ is the gas metallicity computed by the total mass of atomic metals in gas phase.  

For gas at solar metallicity with $n_{c} = \rm 10^3 cm^{-3}$ and $T_{\rm c} = \rm 50 \, K$, the accretion timescale becomes $\rm  \tau_{\rm gg,0} =  2.0 \,Myr$ (see \citealt{2013MNRAS.432..637A}). \citet{2014MNRAS.445.3039D} show that  this value reproduces the observed dust-to-gas ratio of local galaxies over a wide range of gas metallicity; it also  provides predictions consistent with the upper limits inferred from deep ALMA and PdB observations of galaxies at $z > 6$ \citep{2015MNRAS.451L..70M}.

\texttt{dustyGadget} computes $\tau_{\rm gg}$ in the cold phase of star-forming particles by relying on the physical
conditions in the model, unlike the usual schemes which assume fixed values for $n_c$ and $T_c$ (see e.g.  \citet{2015MNRAS.451L..70M,2016MNRAS.462.3130M}). $Z_c$ is consistently computed from the mass of atomic metals and gas available in the model. 
In agreement with the ISM scheme (see Appendix \ref{sec:AppAISMModel}) we compute $n_c$ by assuming that it is entirely composed of a neutral atomic gas mixture of hydrogen and helium with a mean molecular weight $\mu = 1.22$ \citep{2001PhR...349..125B}. In reality, the cold ISM phase observed in galaxies comprises both atomic and  molecular regions and a more realistic model should rely on a consistent implementation of both the H$_2$ formation process on dust grains and its photo-dissociation under a Lyman-Werner flux. We defer the treatment of these additional mechanisms to a future study, also linking star formation to H$_2$, rather than total cold gas including HI.

We also note that in Equation 6 the value of the scaling factor $\tau_{\rm gg,0}$ depends on the evolution of $T_{\rm c}$. As summarized in Appendix \ref{sec:AppAISMModel} and detailed in \citet{2003MNRAS.339..289S}, the multi-phase implementation of the ISM does not explicitly follow the evolution of $T_{\rm c}$ but only assumes a fiducial, average value of $T_{\rm c} = 10^3$~K, i.e. a constant energy per unit mass of the cold gas ($u_c$).  While the evolution of the hot phase is proven not to depend strongly on this assumption (see references in Appendix \ref{sec:AppAISMModel}), grain growth becomes efficient at cooler ambient temperatures (i.e. $T_{\rm c} \approx 50 - 100$~K), but an astrophysical characterization of this environment is still unknown. \citet{2018MNRAS.476.1371C} have shown that in cold molecular clouds ($n \approx 1000$~cm$^{-3}$, $T \approx 10-20$~K) dust grains can easily develop icy mantles so that  their growth has a problematic chemical justification. In the cold neutral medium  ($n \approx 30$~cm$^{-3}$, $T \approx 100$~K) grains can probably  grow, particularly if Coulomb focusing enhances the collision rate, as suggested by \citet{weingartner1999} and \citet{ zhukovska2018}. Hereafter, as a compromise, a value of $T_{\rm c} = 50$~K is adopted in our model. 

As for the destruction term described in \ref{sec:dustDestruction}, the numerical implementation of the grain growth process is straightforward in our SPH scheme: at each time step we compute the fraction of cold mass of star forming SPH particles, account for the dust mass in the cold phase and finally increase it by the growth term. 

\subsubsection{Grain sputtering}

Once the grains enter a hot plasma ($T_{\rm h} \gtrsim 10^6$~K) they are sputtered away by thermal collisions with both protons and helium nuclei. This process has been modeled in the past by many authors \citep{1994ApJ...431..321T, 1979ApJ...231...77D, 1979ApJ...231..438D, 1987ASSL..134..491S}  and included in models of dust evolution in elliptical galaxies, where the hot phase largely dominates the galactic ISM \citep{1995ApJ...448...84T}. The sputtering timescale $\tau_{\rm sp}$ on spherically modeled grains depends on the plasma number density $n_{\rm h}$, the temperature $T_{\rm h}$ and the grain size $a(t)$. In the above models it is defined as inversely proportional to the rate at which $a$ decreases in time, i.e.:
\begin{equation*}
\label{eq:tausput}
\begin{array}{lcl} \tau_{\rm sp} & = & a \left|\frac{da}{dt}\right|^{-1} \\ \left|\frac{da}{dt}\right| & = & -3.2\times10^{-18} {\rm cm}^4{\rm s}^{-1} \left(\frac{\rho}{m_{\rm p}}\right)\left[\left(\frac{2\times10^{6}K}{T_{\rm h}}\right)^{2.5}+1\right]^{-1}, 
\end{array}
\end{equation*}
\noindent
where $\rho$ and $m_p$ are the gas density and proton mass respectively.
Also note that this approximation is valid for both silicate and carbon dust. Coherently with the grain growth assumptions and the temperature of the hot phase of SPH particles, here we adopt an explicit expression for $\tau_{\rm sp}$ by assuming spherical grains with typical size of $a \approx 0.1$~$\mu {\rm m}$ and collisional ionization of the gas. This leads to the formula:
\begin{equation}
\tau_{\rm sp}=	 1.68 \times 10^{-4} {\rm Gyr} \left(\frac{n_{\rm h}}{\rm cm^{-3}}\right)^{-1} \left[\left(\frac{2\times10^{6}}{T_{\rm h}}\right)^{2.5}+1\right];
 \label{tausputFin}
 \end{equation}
 \noindent
in this way the sputtering term in Equation \ref{eq:dustEvolution} becomes $-M_{\rm d}/(\tau_{\rm sp}/3)$.
Finally note that for temperatures $T_{\rm h} < 10^6$~K, sputtering becomes very inefficient and dust grains could easily survive in a diffuse, photo-ionized IGM once spread by galactic winds. 

\subsection{Spreading of atomic metals and dust by galactic winds}
\label{sec:dustInWinds}
In its first implementation \texttt{dustyGadget} adopts the wind prescription implemented in \citet{2003MNRAS.339..289S}, on top of which metals are spread in the surrounding gas as described by \citet{2007MNRAS.382.1050T}. At the end of stellar evolution, metals and dust are then distributed in the surroundings of a star forming region by using a spline-type kernel of the SPH scheme and by weighting over 64 neighbours according to the influence region of each particle. The dust distribution simply follows the atomic metal spreading without accounting for any momentum transfer through dust grains (see \citealt{mckinnon2018} for a recent implementation that accounts for dynamical forces acting on dust particles). At the same time, dusty particles associated with galactic winds evolve in their hot phase through sputtering. As a result, the dust-to-metal ratio will be modulated
depending on the environment, attaining different values for the galactic ISM, CGM and IGM.

\begin{table*}  
\begin{tabular} {|l|l|l|l|l|l|l|l} \hline

Name & Type & Simulation & ISM & Yields & Accretion  & Thermal Sputtering & Destruction \\
\hline
Popping+17           & semi-analytic     & "Fiducial"    & $b$ & $f$         & $q,f$ & $r$ & $v$\\  
Mancini+15           & semi-numeric      & "SN+AGB+GG"   & $a$ & $g,i$       & $p$   & -   & $s$ \\
Gioannini+17         & analytic          & "Alternative" & $c$ & $h$         & $z,t$   & -   & $t$ \\
Aoyama+18            & SPH-GADGET3-OSAKA & "L50n512"     & $d$ & $l,u$       & $p$   & $r$ & $s$\\
McKinnon+17          & MovingMesh-AREPO  & "L25n512"     & $e$ & $f$         & $f$   & $r$ & $s$  \\
\texttt{dustyGadget} & SPH-Gadget2       & "RefRun"      & $a$ & $g,i,m,n,o$ & $p$   & $r$ & $s$\\
\hline
\hline
\end{tabular}
\caption{Summary of theoretical models adopted for a comparison with \texttt{dustyGadget}. 
References for ISM models: $a$: \citet{2003MNRAS.339..289S,2010MNRAS.407.1003M}, $b$: \citet{2010MNRAS.402..173A,2015MNRAS.453.4337S}, $c$:  \citet{2017MNRAS.471.4615G,2017MNRAS.464..985G}, $d$: \citet{2017MNRAS.466..105A}, $e$: \citet{2013MNRAS.436.3031V}. References for stellar dust yields and dust growth models: $f$:  \citet{1998ApJ...501..643D,2016ApJ...825..136D}, $g$: \citet{2008A&A...479..453Z}, $h$: \citet{2011arXiv1107.4541P}, $i$:  \citet{2007MNRAS.378..973B}, $l$: \citet{2011EP&S...63.1027I,2012MNRAS.424L..34K,2011MNRAS.415.2920I}, $m$:  \citet{2019MNRAS.484.2587M}, $n$: \citet{2017MNRAS.467.4431D}, $o$: \citet{2016A&A...587A.157B}, $p$: \citet{Hirashita2014}, $q$: \citet{2014A&A...562A..76Z}, $u$: \citet{2017AJ....153...85S}, $v$: \citet{2015ApJ...803....7S}, $z$: \citet{2000PASJ...52..585H}. References for grain sputtering: $r$: \citet{1995ApJ...448...84T}. Grain destruction references: $s$: \citet{McKee1989}, $t$: \citet{2013MNRAS.432..637A}}
\label{tab:models}
\end{table*}

\section{Galaxy formation simulation}
\label{sec:galForm}
The features of \texttt{dustyGadget} have been exploited in a new hydrodynamical simulation performed on a periodic, comoving box size of $30 \,h^{-1}\rm  cMpc$, and assuming a $\Lambda$CDM cosmology consistent with WMAP7 data release \citep{2011ApJS..192...18K}\footnote{$H_{0}= 100 \, h \, \textrm{km}\,\textrm{s}^{-1}\,\textrm{Mpc}^{-1}$ with $h = 0.7$ ,$\Omega_{\rm m,0}=0.3$, $\Omega_{\rm b,0}=0.04$, $\Omega_{\Lambda,0}=0.7$, $\Omega_{\rm tot,0}=1.0$, $\sigma_{8}=0.9$}. The simulation starts from a neutral gas configuration at $z=100$ and zero metallicity, and evolves $\approx 320^{3}$ particles per gas and dark matter (DM) species with masses of $9\times 10^6 h^{-1} M_\odot$ and $ 6 \times 10^7 h^{-1} M_\odot$  respectively, down to $z = 4$; 30 outputs at intermediate redshifts are stored during the run. For a better comparison with \citet{2015MNRAS.451L..70M}, we adopted a chemical set-up close to the one described in \citet{2010MNRAS.407.1003M} including molecules and atomic metals, while simulating a larger cosmic volume to match the galaxy sample in \citet{2016MNRAS.462.3130M}. Hereafter, we briefly summarize the relevant properties of the run. Stars are formed from the cold gas phase once the density exceeds a threshold value of $n_{\rm th} = 132 \, h^{-2} \rm cm^{-3}$ (physical); this choice allows the capture of all the relevant phases of cooling until the onset of runaway collapse \citep{2009A&A...503...25M}. As in \citet{2007MNRAS.382..945T}, stellar populations follow an initial mass function (IMF) consistent with the metallicity of  stellar particles ($Z_\star$) and the transition from Pop\ III to Pop\ II stars is modeled by assuming that metal-fine structure cooling is efficient at a gas critical metallicity $Z_{\rm cr}=10^{-4}Z_{\odot}$ \citep{2007MNRAS.379..963M}. Below $Z_{\rm cr}$, the IMF is assumed to follow a Salpeter power-law slope in the mass range $[100, 500]~M_{\odot}$, while above  $Z_{\rm cr}$ a standard Salpeter IMF in the mass range  $[0.1,100]~M_{\odot}$ is adopted. 
An extensive investigation on the impact of the adopted Pop\ III IMF and $Z_{\rm cr}$ on the earliest phases of star formation and 
chemical enrichment would require a higher mass resolution that can be achieved only by simulating smaller cosmological volumes \citep{xu2016}. 
We will defer this point to a future study where an approximate treatment of radiative feedback will be also implemented in our model \citep{maio2016}.

Galactic-scale winds associated with star-forming regions are assumed with a constant velocity of $500~\textrm{km}\,\textrm{s}^{-1}$, in line with recent estimates of star-formation driven outflows in normal galaxies at $z > 4$ \citep{2019ApJ...886...29S,2020A&A...633A..90G}. Finally, radiative feedback is implemented by adopting a cosmic UV background \citep{1996ApJ...461...20H} and accounting for photo-ionisation, which affects gas cooling and hence star formation. Although this chemical evolution model can be extended to track a large number of metal species, in the current simulation we restrict the analysis to the following atomic metals: C, O, Mg, S, Si and Fe. 
 
Dark matter halos and their sub-structures are found by running the halo finder code \texttt{AMIGA} \citep{2004MNRAS.351..399G, 2009ApJS..182..608K} and the catalogue has been verified to be consistent with the friends-of-friends (FOF) and SUBFIND implemented in \texttt{Gadget} \citep{2001MNRAS.328..726S} and adopted in \citet{2016MNRAS.462.3130M}. Galaxies are identified as bound groups of at least 32 total (DM+gas+star) particles; only galaxies containing  at least 10 stellar particles are considered.

\begin{table*}
\begin{tabular} {|l|c|c|c|c|c} 
\hline
Name    	    & z      & Log$(M_{\star})$   & $\rm SFR (UV;IR)$        & Log$(M_{\rm d})$      & $T_{\rm d}$  \\ 
                &        & $[M_\odot]$           & $[M_\odot \rm yr^{-1}]$  & $[M_\odot]$        & $\rm [K]$  \\ 
\hline
A2744\_YDA$^a$  & 8.382  & $9.29\substack{+0.25 \\ -0.17}$ & $20\substack{+17.6 \\ -9.5}$ & $6.74\substack{+0.66 \\ -0.16}$ & 37-63  
\\
\addlinespace[0.05cm]
MACS0416\_Y1$^b$ & 8.312 & $8.38\substack{+0.11 \\ -0.02}$ & $55\substack{+175 \\ -0.2}$ & $6.91\substack{+0.07 \\ -0.09}$ ; $6.56\substack{+0.07 \\ -0.1}$ & 40 ; 50
\\
\addlinespace[0.05cm]
MACS0416\_Y1$^0$ & 8.312 & $8.38\substack{+0.11 \\ -0.02}$ & - & $6.23$ ; $5.76$ & 50 ; 90
\\

\addlinespace[0.05cm]
z8-GND-5296$^{c}$ & 7.508 & $9.7\pm 0.3$ & $23.4$ ; $<113$ & $<8.69$ ; $<8.13$ ; $<7.81$ & 25 ; 35 ; 45              
\\
\hline
A1689-zD1$^{d}$ & 7.500 & $9.4\pm 0.1$ & $14.0\substack{+8.0 \\ -8.0}$  & $7.2\pm 0.2$  & 40.5 
\\
\addlinespace[0.05cm]
B14-65666$^e$    & 7.170 & $8.89\substack{+0.01 \\ -0.04}$ & $200\substack{+82 \\ -39}$ & $7.05\substack{+0.04 \\ -0.09}$ ; $6.97\substack{+0.03 \\ -0.09}$ ; $6.91\substack{+0.08 \\ -0.1}$ & 48 ; 54 ; 61               
\\
\addlinespace[0.05cm]
BDF-3299$^{f}$ & 7.109 & $9.30 \pm 0.30$ & $ 5.7 ; -$ & $< 7.32$ ; $< 6.50$ & 27.6 ; 45              
\\
\addlinespace[0.05cm]
BDF-512$^{f}$  & 7.008 & $9.30 \pm 0.30$ & 6.0 ; -& $<7.72$ ; $<6.89$ & 27.6 ; 45              
\\
\addlinespace[0.05cm]
IOK-1$^g$        & 6.960 & $9.70 \pm 0.30$ & $20.4 ; < 16.3$  & $<7.84$ $<7.29$ ; $<6.98$ & 25 ; 35 ; 45           
\\
\addlinespace[0.05cm]
SPT0311-58E$^m$  & 6.900 & $10.54\substack{+0.15 \\ -0.24}$ & $13.0 ; < 540 \pm 175$  & $8.60\substack{+0.18 \\ -0.30}$ & 36 - 115           
\\
\addlinespace[0.05cm]
SDF-46975$^{f}$  & 6.844 & $9.80 \pm 0.30$ & 15.4 ; - & $<7.76$ ; $<6.94$ & 27.6 ; 45              
\\
\addlinespace[0.05cm]
A1703-zD1$^g$    & 6.800 & $9.20 \pm 0.30$ & $9.0$ ; $13.8$ & $<7.76$ ; $<7.22$ ; $<6.91$ & 25 ; 35 ; 45                  
\\
\addlinespace[0.05cm]
Himiko$^g$       & 6.595 & $9.80 \pm 0.30$ & $32.3$ ; $11.4$ & $<7.67$ ; $<7.13$ ; $<6.83$ & 25 ; 35 ; 45               
\\
\addlinespace[0.05cm]
HCM 6A$^g$       & 6.560 & $9.50 \pm 0.30$ & $13.7$ ; $24.5$ & $<8.0$ ; $<7.47$ ; $<7.17$ & 25 ; 35 ; 45                 
\\
\hline
HFLS 3$^l$ & 6.34 & $10.70$ & $-;1320$ & $8.48$ & $24-50$
\\
\addlinespace[0.05cm]
ALESS061.1$^h$ & 6.120 & $10.59\substack{+0.02 \\ -0.25}$ & $1380.38\substack{+525.08 \\ -690.55}$ & $8.579\substack{+0.62 \\ -0.22}$ & $57.0\substack{+0.62 \\ -0.22}$
\\
\addlinespace[0.05cm]
ALESS072.1$^h$ & 5.820 & $10.95\substack{+0.14 \\ -0.41}$ & $549.54\substack{+863.00 \\ -414.64}$ & $8.930\substack{+0.61 \\ -0.44}$ & $44.0\substack{+18 \\ -14}$
\\
\addlinespace[0.05cm]
HZ1$^i$ & 5.690 & $10.00 \pm 0.30$ & $31.7$; - & $<7.93$ ; $<7.42$ ; $<7.13$ & 25 ; 35 ; 45
\\
\addlinespace[0.05cm]
HZ2$^i$ & 5.670 & $10.00 \pm 0.30$ & $32.4$; - & $<7.91$ ; $<7.40$ ; $<7.11$ & 25 ; 35 ; 45
\\
\addlinespace[0.05cm]
HZ10$^i$ & 5.660 & $10.30 \pm 0.30$ & $50.12$; - & $<9.07$ ; $<8.56$ ; $<8.27$ & 25 ; 35 ; 45
\\
\addlinespace[0.05cm]
HZ3$^i$ & 5.550 & $9.89 \pm 0.30$ & $23.5$; - & $<8.14$ ; $<7.64$ ; $<7.34$ & 25 ; 35 ; 45
\\
\addlinespace[0.05cm]
HZ9$^i$ & 5.550 & $9.79 \pm 0.30$ & $23.5$; - & $<8.66$ ; $<8.16$ ; $<7.87$ & 25 ; 35 ; 45
\\
\addlinespace[0.05cm]
ALESS065.1$^h$ & 5.680 & $10.74\substack{+0.17 \\ -0.48}$ & $436.51\substack{+765.75 \\ -321.69}$ & $8.910\substack{+0.56 \\ -0.48}$ & $44.0\substack{+17 \\ -16}$
\\
\addlinespace[0.05cm]
\hline
HZ4$^i$ & 5.540 & $10.20 \pm 0.30 $ & $40.8$; - & $<8.25$ ; $<7.75$ ; $<7.46$ & $25$ ; $35$ ; $45$
\\
\addlinespace[0.05cm]
HZ5$^i$ & 5.300 & $10.40 \pm 0.30$ & $64.7$; - & $<7.90$ ; $<7.40$ ; $<7.11$ & $25$ ; $35$ ; $45$
\\
\addlinespace[0.05cm]
HZ6$^i$ & 5.290 & $10.20 \pm 0.30$ & $48.0$; - & $<8.02$ ; $<7.52$ ; $<7.23$ & $25$ ; $35$ ; $45$
\\
\addlinespace[0.05cm]
HZ7$^i$ & 5.250 & $9.96 \pm 0.30$ & $27.0$; - & $<7.94$ ; $<7.44$ ; $<7.15$ & $25$ ; $35$ ; $45$
\\
\addlinespace[0.05cm]
ALESS001.2$^h$ & 5.220 & $10.90\substack{+0.13 \\ -0.39}$ & $416.87\substack{+758.03 \\ -339.24}$ & $8.69\substack{+0.61 \\ -0.37}$ & $45.0\substack{+21 \\ -16}$
\\
\addlinespace[0.05cm]
HZ8$^i$ & 5.140 & $9.89 \pm 0.30$ & $23.5$; - & $<7.84$ ; $<7.34$ ; $<7.06$ & $25$ ; $35$ ; $45$
\\
\addlinespace[0.05cm]
ALESS001.1$^h$ & 4.780 & $10.97\substack{+0.15 \\ -0.42}$ & $602.56\substack{+1259.53 \\ -432.74}$ & $9.12\substack{+0.49 \\ -0.46}$ & $42.0\substack{+20 \\ -14}$
\\
\addlinespace[0.05cm]
ALESS073.1$^h$ & 4.780 & $10.64\substack{+0.05 \\ -0.22}$ & $794.33\substack{+790.56 \\ -447.59}$ & $8.97\substack{+0.34 \\ -0.32}$ & $46.0\substack{+17 \\ -14}$
\\
\addlinespace[0.05cm]
ALESS055.2$^h$ & 4.680 & $10.42\substack{+0.22 \\ -0.55}$ & $229.10\substack{+462.73 \\ -168.84}$ & $8.61\substack{+0.56 \\ -0.42}$ & $44.0\substack{+18 \\ -16}$
\\
\addlinespace[0.05cm]
ALESS069.3$^h$ & 4.680 & $10.36\substack{+0.22 \\ -0.55}$ & $194.98\substack{+421.61 \\ -144.86}$ & $8.55\substack{+0.56 \\ -0.41}$ & $44.0\substack{+18 \\ -16}$
\\
\addlinespace[0.05cm]
ALESS087.3$^h$ & 4.680 & $10.43\substack{+0.22 \\ -0.54}$ & $234.42\substack{+490.01 \\ -169.86}$ & $8.63\substack{+0.56 \\ -0.42}$ & $44.0\substack{+18 \\ -16}$
\\
\addlinespace[0.05cm]
ALESS099.1$^h$ & 4.620 & $10.36\substack{+0.23 \\ -0.55}$ & $199.53\substack{+417.07 \\ -145.82}$ & $8.55\substack{+0.56 \\ -0.48}$ & $44.0\substack{+18 \\ -16}$
\\
\addlinespace[0.05cm]
ALESS035.2$^h$ & 4.570 & $10.21\substack{+0.23 \\ -0.55}$ & $131.83\substack{+294.75 \\ -99.47}$ & $8.40\substack{+0.55 \\ -0.42}$ & $43.0\substack{+19 \\ -15}$
\\
\addlinespace[0.05cm]
ALESS103.3$^h$ & 4.570 & $10.21\substack{+0.22 \\ -0.57}$ & $131.83\substack{+304.69 \\ -100.20}$ & $8.40\substack{+0.56 \\ -0.42}$ & $43.0\substack{+19 \\ -15}$
\\
\addlinespace[0.05cm]
\hline
ALESS069.2$^h$ & 4.380 & $10.38\substack{+0.22 \\ -0.52}$ & $208.93\substack{+467.15 \\ -152.69}$ & $8.64\substack{+0.52 \\ -0.44}$ & $43.0\substack{+18 \\ -15}$
\\
\addlinespace[0.05cm]
ALESS088.2$^h$ & 4.280 & $10.64\substack{+0.14 \\ -0.40}$ & $151.36\substack{+285.16 \\ -114.20}$ & $8.58\substack{+0.53 \\ -0.40}$ & $40.0\substack{+20 \\ -13}$
\\
\addlinespace[0.05cm]
ALESS023.1$^h$ & 4.070 & $11.18\substack{+0.20 \\ -0.54}$ & $891.25\substack{+1150.49 \\ -575.02}$ & $8.90\substack{+0.35 \\ -0.25}$ & $49.0\substack{+17 \\ -15}$
\\
\addlinespace[0.05cm]
ALESS076.1$^h$ & 3.970 & $11.01\substack{+0.20 \\ -0.49}$ & $691.83\substack{+929.98 \\ -451.95}$ & $8.90\substack{+0.40 \\ -0.25}$ & $44.0\substack{+17 \\ -13}$
\\
\addlinespace[0.05cm]
ALESS037.2$^h$ & 3.830 & $10.55\substack{+0.10 \\ -0.27}$ & $213.80\substack{+562.45 \\ -}$ & $8.34\substack{+0.52 \\ -0.47}$ & $47.0\substack{+19 \\ -16}$
\\
\addlinespace[0.05cm]
ALESS002.2$^h$ & 3.780 & $11.00\substack{+0.19 \\ -0.51}$ & $588.84\substack{+996.05 \\ -393.86}$ & $8.64\substack{+0.35 \\ -0.22}$ & $48.0\substack{+18 \\ -14}$
\\
\addlinespace[0.05cm]
ALESS068.1$^h$ & 3.780 & $10.97\substack{+0.19 \\ -0.57}$ & $407.38\substack{+714.64 \\ -287.15}$ & $8.67\substack{+0.40 \\ -0.34}$ & $47.0\substack{+17 \\ -15}$
\\
\addlinespace[0.05cm]
MACSJ0032-arc$^n$ & 3.631 & $7.89\pm 0.04$ & $51\substack{+7 \\ -10}$ & $5.49\substack{+0.14 \\ -0.20}$ & $43.0\pm 5$
\\
\addlinespace[0.05cm]
ALESS110.5$^h$ & 3.620 & $10.55\substack{+0.16 \\ -0.65}$ & $151.36\substack{+451.20 \\ -147.38}$ & $8.61\substack{+0.55 \\ -0.95}$ & $42.0\substack{+18 \\ -16}$
\\
\addlinespace[0.05cm]
ALESS110.1$^h$ & 3.580 & $11.05\substack{+0.24 \\ -0.51}$ & $891.25\substack{+339.02 \\ -401.47}$ & $8.71\substack{+0.35 \\ -0.26}$ & $66.0\substack{+1 \\ -15}$
\\
\addlinespace[0.05cm]
ALESS116.1$^h$ & 3.580 & $10.92\substack{+0.17 \\ -0.44}$ & $549.54\substack{+652.72 \\ -330.76}$ & $8.52\substack{+0.17 \\ -0.27}$ & $48.0\substack{+16 \\ -12}$
\\
\addlinespace[0.05cm]
ALESS116.2$^h$ & 3.580 & $11.19\substack{+0.16 \\ -0.50}$ & $467.73\substack{+423.52 \\ -293.96}$; - & $8.60\substack{+0.30 \\ -0.16}$ & $44.0\substack{+16 \\ -10}$
\\
\hline
\hline
\end{tabular}
\caption{Physical properties of the galaxy sample in Figure \ref{fig:Fig3MdMstar} collected from the literature: galaxy redshift $z$, logarithm of the stellar mass $M_{\star}$, star formation rate (when available the values derived from the UV and IR components are indicated), logarithm of the dust mass $M_{\rm d}$ and dust temperature. When the dust mass are derived from scaling relations by assuming a value of $T_{\rm d}$, the corresponding values are listed in the same order in both columns.  References: $a$: \citet{2017ApJ...837L..21L}; $b$: \citet{2019ApJ...874...27T}; $c$: \citet{2015A&A...574A..19S}; $d$:  \citet{2017MNRAS.466..138K}; $e$: \citet{2018arXiv180600486H}; $f$: \citet{2015MNRAS.452...54M,2015MNRAS.451L..70M}; $g$: \citet{2006Natur.443..186I}; $h$: \citet{2015ApJ...806..110D}; $i$: \citet{2015Natur.522..455C,2016MNRAS.462.3130M,2017ApJ...847...21F}; $l$: \citet{2014ApJ...790...40C}; $m$: \citet{2017ApJ...842L..15S,2018Natur.553...51M}; $n$: \citet{2017A&A...605A..81D}; $o$: \citet{Bakx2020}.}
\label{tab:ObsTable}
\end{table*}

\section{Reference theoretical models and observations}
\label{sec:ThModels}

To  assess the reliability of our results, \texttt{dustyGadget} is compared with a series of analytic, semi-analytic and numerical schemes. We selected the study of \citet{2017MNRAS.471.4615G}, which combines a series of analytic prescriptions to evolve the number density of a pre-assigned class of galaxy morphologies, and the semi-analytic model of \citet{2017MNRAS.471.3152P}; both models include processes of dust formation and evolution. Our previous study, introduced in \citet{2015MNRAS.451L..70M}, is added as semi-numerical model and complemented with two reference hydrodynamical schemes: a moving-mesh-based implementation in \texttt{AREPO} \citep{2017MNRAS.468.1505M} and a SPH, \texttt{Gadget}-based code \citep{aoyama2018}.

Table \ref{tab:models} summarizes the literature references with the algorithm type, the ISM model, the implementation of dust yields and the physics of dust evolution: grain growth and destruction\footnote{Other important processes, as for example the coagulation of dust grains, have not been considered because they do not change the total mass of dust.}. Additional details and a summary of their efficiency parameters are provided in Appendix \ref{sec:APPB_ThModels}.

The production of dust by stellar sources relies on different yields and physical assumptions: recent models implement mass and metallicity dependent tabulated values and account for the partial destruction of SN dust by the reverse shock (\citealt{2015MNRAS.451L..70M}, \texttt{dustyGadget}), while other implementations still rely on extrapolated trends, including the RS effects through an average correction (e.g. \citealt{2017MNRAS.471.3152P}). A disagreement on the sources of dust production is also present: \citet{2017MNRAS.471.3152P}, \citet{2017MNRAS.468.1505M}, \citet{aoyama2018},  assume that SN\ Ia can produce dust, while the remaining models are more conservative and exclude these supernovae as dust producers. Finally, \texttt{dustyGadget} is the only code that explicitly accounts for the contribution of PISN at the highest redshifts (see Section 2.2 for more details).

The implementation of dust evolution requires a two-phase description of the galactic ISM in all approaches, either by assuming a certain cold fraction $x_c$ \citep{2015MNRAS.451L..70M, 2017MNRAS.471.4615G, 2017MNRAS.468.1505M, aoyama2018}, or by consistently taking it from star forming particles (\texttt{dustyGadget}) or by deriving it from an explicit description of the H$_2$ molecular phase  \citep{2017MNRAS.471.3152P}. Differences also exist in the implementation of the accretion and destruction mechanisms and in the adopted time scales. Table \ref{tab:modelValues} shows that $\tau_{\rm gg,0}$ varies in the range $[1.2-4.0]$~Myr across models, while the value of the swept mass and destruction efficiencies are tuned either by assuming different reference values for the shock speed or by correcting the supernova rates across supernova types (see the values of $f_{\rm SN}$ and $\epsilon_d$ and M$_s$ in Table \ref{tab:modelValues}). Finally note that only \citet{2017MNRAS.471.3152P} distinguish between carbonaceous- and silicon-based dust when destroying the grains, while \citet{aoyama2018} is the only model that considers a grain size distribution. 

The implementation of dust sputtering, when present, is consistent across models and relies on the description of \citet{1995ApJ...448...84T}.\newline \newline

To compare the predictions of our simulation with observations available at $ z \ge 4$, we collected a sample of dusty galaxies in the redshift range $3.5 < z < 9.5$. Table \ref{tab:ObsTable} summarizes their physical properties: redshift ($z$), Log$(M_{\star})$, SFR (derived from the UV or the IR flux), Log$(M_{\rm d})$ and the dust temperature $T_{\rm d}$\footnote{Note that when the dust mass depends on different assumptions for $T_{\rm d}$, the values of temperature and mass are listed in the same order.}. The last two values are either inferred by dust continuum detections or derived as in \citet{2015A&A...574A..19S} and \citet{2015MNRAS.451L..70M}. Galaxies are also grouped in redshift bins of $\Delta z = 1$, centered at $z = 9,8,7,6,5,4$, and are selected in the stellar mass range covered by our simulated sample: $8.0 \leq {\rm Log(}M_{\star}/M_{\odot}) \leq 11.0$. All the galaxies have a spectroscopically confirmed redshift, by a detection of either  the Lyman-$\alpha$ line or some metal lines (e.g. [C\ II] or [O\ III]), while data taken from the ALMA Survey of Sub-millimeter Galaxies in the Extended Chandra Deep Field South (ALESS, \citealt{2015ApJ...806..110D}) has photometric redshifts, constrained by a sub-sample of spectroscopic observations (see the original references for more details). 

At $z > 6.5$ our galaxy sample comprises single observations of normal star forming galaxies with $ \rm SFR \leq 50$~$M_{\odot}$ yr$^{-1}$ with the only exception of B14-65666, which has an inferred $\rm SFR$ of $ \approx 200$~$M_{\odot}$ yr$^{-1}$. 
We also mix normal star forming galaxies of \citet{2015Natur.522..455C}, having $\rm SFR \in [23.5, 64.7]$~$M_{\odot}$ yr$^{-1}$, with the sample of the ALESS survey, where star formation rates range from $\rm SFR \geq 100$~$M_{\odot}$ yr$^{-1}$ up to $\rm SFR \approx 1400$~$M_{\odot}$ yr$^{-1}$ (i.e. see the ALESS061.1 galaxy at $z=6$). 

At $z \la 6$, galaxies with $\rm SFR \geq 100$~$M_{\odot}$ yr$^{-1}$ have ${\rm Log(}M_{\star}/M_{\odot}) > 10.25$ and their dust masses are typically one order of magnitude larger than the values of normal star forming galaxies. It should be noted, on the other hand, that even when a direct detection of the continuum flux is available, the inferred values of $M_{\rm d}$ depend on the adopted dust temperature $T_{\rm d}$ and emissivity\footnote{Here we point out that the dependence $M_{\rm d} ( T_{\rm d})$ applies to masses inferred in observations while the dust mass computed in our simulation does not require any assumptions on its temperature.}. The case of MACS0416\_Y1 provides a good example of a spectroscopically confirmed galaxy at $z\approx 8.3$ with a direct detection of the continuum flux, but the estimated dust mass varies by a factor 2.2 when the assumed dust temperature differs by $\Delta T_d \approx10$~K. Whenever the dust continuum emission is not detected, the table reports upper limits on $M_{\rm d}$ for the samples in \citet{2015MNRAS.452...54M} and \citet{ 2015Natur.522..455C}; these are derived from relations introduced in \citet{2015A&A...574A..19S} and complemented by the computations in \citet{2015MNRAS.451L..70M,2016MNRAS.462.3130M}. Also in this case, we emphasize the tight dependence of these upper limits on the assumed grain temperatures $T_{\rm d} \in [25, 45]$~K. 

\section{Results}
\label{sec:results}

In this section we discuss the results of our reference simulation (RefRun), which accounts for the full set of physical processes implemented in \texttt{dustyGadget}. The corresponding set of physical parameters are listed in Table \ref{tab:modelValues}. As a comparison, we also explore a case where we do not consider grain growth in the ISM, i.e. we assume $\tau^{-1}_{\rm gg,0}=0$. We refer to this run as "ProdOnly", to indicate that dust is produced only by stellar sources. 

Section \ref{sec:dustZ} shows the redshift evolution of the cosmic density parameter $\Omega_{\rm d}$ while Section \ref{sec:galZoo} focuses on statistical properties of the dusty galaxy sample found in the RefRun: the dust mass function (\ref{sec:MF}), the dust-to-stellar mass relation (\ref{sec:DSM}) and finally the dust-to-gas and dust-to-metal ratios (\ref{sec:Ratios}). A qualitative analysis of the dusty environment of a massive representative halo found at $z\approx4$ is finally provided in Section \ref{sec:dustyMMHalo}.

\begin{figure}
\centering
\includegraphics[angle=-90,width=0.45\textwidth, trim={1.15cm 0.1cm 0 0},clip]{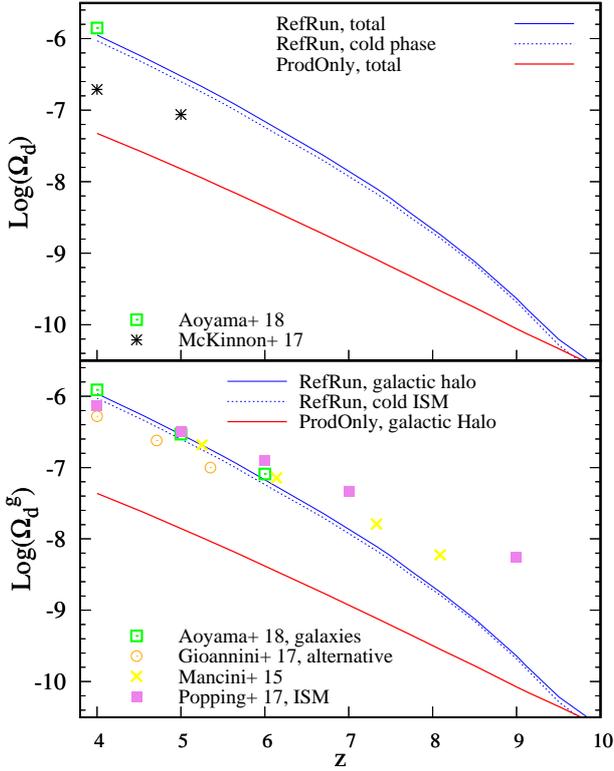}
\caption{{\bf Top panel}: logarithm of $\Omega_{\rm d}(z) \equiv \rho_d (z) / \rho_{c,0}$ as function of redshift $z$; $\rho_{c,0} = 2.775\times 10^{11} h^2$~$M_{\odot}$~Mpc$^{-3}$. The solid blue line refers to $\rm Log(\Omega_d)$ computed from the reference run including both dust production and evolution (RefRun, total), while the blue dashed line shows values deriving from the cold phase environment in RefRun. $\rm Log(\Omega_d)$, computed from a run in which dust is produced only by stellar sources (prodOnly, total), is shown as solid red line. Black asterisks are values extrapolated from \citet{2017MNRAS.468.1505M} (see their Fig. 4), while  green squares refer to \citet{aoyama2018} (Fig. 4). {\bf Bottom panel}: same as in the top panel but evaluated from the dust mass contained in galactic halos ($\Omega^g_{\rm d}(z)$); here the dashed blue line refers the cold ISM of the halos. Green empty-squares show data taken from \citet{aoyama2018} (Fig. 5, top panel), violet filled-squares are taken from \citet{2017MNRAS.471.3152P} (Fig. 10, "fiducial" run), gold  empty-circles show the "alternative scenario" in Fig. 6 of \citet{2017MNRAS.471.4615G} and finally, gold crosses refer to data in \citet{2015MNRAS.451L..70M}.}
\label{fig:FigOmegad}
\end{figure}
 
\subsection{Cosmic dust density parameter}
\label{sec:dustZ}
Here we investigate the redshift evolution of the cosmic dust density parameter $\Omega_{\rm d}$ in the redshift range $4 \leq z \leq 10$. $\Omega_{\rm d} (z)$ is defined as $\Omega_{\rm d} (z) \equiv \rho_{\rm d} (z) / \rho_c,0$, where $\rho_{\rm d}(z)$ is the density of cosmic dust in the cosmological volume and $\rho_c,0$ is the critical density of the universe at $z=0$\footnote{In accordance with the WMAP-7 cosmology adopted in our simulation the cosmic critical density at $z=0$ is $\rho_{\rm c,0} \approx 2.775\times 10^{11}$~M$_{\odot} h^2$ cMpc$^{-3}$.}. 

The top panel of Figure \ref{fig:FigOmegad} shows $\Omega_{\rm d}(z)$ computed in the RefRun (solid blue line) and in the ProdOnly simulations (solid red line). The dotted blue line shows the same quantity evaluated from the dust mass in the cold phase of the RefRun. Green empty squares and black asterisks are values at $z \geq 4$ taken from \citet{aoyama2018} and \citet{2017MNRAS.468.1505M}, respectively. 

In the redshift interval  $10 \leq z\leq 15$ (not shown in the figure) the blue and red lines overlap, indicating that the dust mass is mainly produced by stellar sources (Pop\ III and Pop\ II stars). An accurate modeling of their stellar yields in this redshift range is required in order to make reliable predictions of the dust mass present in the first galaxies. 
Yet, an extensive investigation of these early phases of metal and dust enrichment would require a higher mass resolution.  To investigate the impact of our setup we performed convergence tests by running three identical hydrodynamical simulations increasing only the number of particles, up to 480$^3$. We found that a convergence between the different runs is obtained at $z \leq 9$. This implies that the dust content of the more massive and evolved galaxies investigated in this paper is not significantly affected by the adopted particle mass.

The increasing difference between the blue and red lines below $z \approx 9$ highlights the importance of 
grain growth in the ISM of the most massive and metal-enriched galaxies \citep{2015MNRAS.451L..70M}, which leads to a cosmic dust density parameter at $z = 4$ that is more than one order of magnitude larger than in the prodOnly case.
The predictions of our reference run are very close to the ones of \citet{aoyama2018} (a similar, \texttt{Gadget}-based implementation), while significantly differing from the values computed by \citet{2017MNRAS.468.1505M}. This is mainly due to the $\tau_{\rm gg,0}$ adopted  in their simulation (see Table \ref{tab:modelValues}). Finally, a comparison between solid and dashed blue lines shows that the largest mass of dust is associated with the cold phase of the ISM, as also found by \citet{aoyama2018} (see their Fig.4). 

The bottom panel shows the evolution of the cosmic dust parameter considering only the dust mass confined in collapsed structures, $\Omega_{\rm d}^{\rm g}(z)$ (i.e. we consider the total mass of dust present in particles belonging to galactic DM halos). 
As expected, the trend is very similar to the one presented in the top panel, since dust is produced by stars and it grows in the galactic ISM.
When compared to other studies, our results show a remarkable agreement with the predictions of \citet{aoyama2018} (green empty squares)
and with semi-analytic/semi-numerical models at $z \leq 6$. At higher $z$, some differences appear. The semi-numerical model of \citet{2015MNRAS.451L..70M} seems to over-produce dust, probably as a result of the more efficient grain growth parametrization adopted, where the grain growth timescale is only modulated by the metallicity of the ISM and it is not sensitive to the cold gas density. 

The comparison  with the results of \citet{2017MNRAS.471.3152P} is complicated by the intrinsic differences among the models. In fact, while their "fiducial" case is the closest to our RefRun in terms of adopted efficiency parameters, it runs on top of a halo catalog generated with an extended Press-Schechter formalism, which converges with predictions of DM-only simulations only on scales larger than 30$h^{-1}$~cMpc. More importantly, 
in their model, grain growth is assumed to occur in molecular clouds, whose number density is inferred from the star formation law \citep{2017MNRAS.471.3152P}. In \texttt{dustyGadget}, dust grains grow in the cold neutral medium, with a number density with a number density inferred from the physical conditions in the ISM (see Appendix \ref{sec:AppAISMModel}).

\subsection{Statistics of dusty galaxies}
\label{sec:galZoo}

This section investigates the redshift evolution of statistical properties and scaling relations of the simulated sample of dusty galaxies. 
The redshift evolution of the dust mass function is discussed in Section \ref{sec:MF}, the relation between stellar mass and dust mass in Section \ref{sec:DSM} and finally the dust-to-gas and dust-to-metal ratios in \ref{sec:Ratios}. 

For an easier comparison with observed values, hereafter all the quantities are physical and converted from  \texttt{Gadget} internal units.

\subsubsection{Dust mass function}
\label{sec:MF}

\begin{figure}
\centering
\includegraphics[angle=-90,width=0.5\textwidth, trim={0.3cm 0cm 0.50cm 0.2cm},clip]{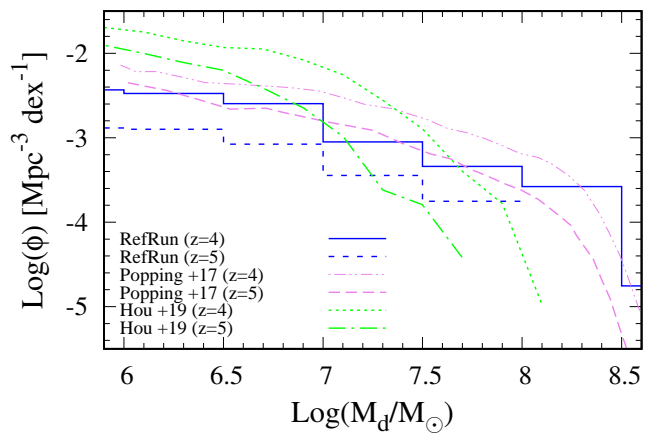}
\caption{Logarithm of the dust mass function $\phi$ of the RefRun as function of Log$(M_{\rm d}/ M_\odot)$ at $z\approx4$ (solid blue line) and $z\approx5$ (dashed blue line). The mass functions at similar redshifts from \citet{2017MNRAS.471.3152P} (violet dashed-double-dotted/long-dashed lines) and \citet{2019MNRAS.485.1727H}, based on the \citet{aoyama2018} simulation (green dotted/dashed-dotted lines) are also shown for a direct comparison.}
\label{fig:FigDMF}
\end{figure}

The dust mass function (DMF, $\phi$) quantifies how galaxies are distributed in dust mass, in the cosmic volume. Figure \ref{fig:FigDMF} shows the DMF\footnote{The mass function $\phi$ is computed with a 0.5 bin size in Log(M$_{\rm d}/ M_\odot$) scale.} of our simulation at redshifts $z=5$ (dashed blue line) and $z=4$ (solid blue line). As expected, the number of galaxies in each dust mass bin grows with time and galaxies progressively populate larger dust mass intervals.  

Our predictions are compared with the results of \citet{2019MNRAS.485.1727H}\footnote{The analysis of \citet{2019MNRAS.485.1727H} is based on the simulation of \citet{aoyama2018}.} (green lines) and of \citet{2017MNRAS.471.3152P} (violet lines). Despite the agreement found in the cosmic dust density parameter (see Figure \ref{fig:FigOmegad}), the amplitude and slope of the DMF show significant variations across models. Compared to \citet{2019MNRAS.485.1727H}, we find a flatter DMF, with fewer galaxies in the low mass bins and more galaxies in the high dust mass intervals. Our slopes are closer to the ones found by \citet{2017MNRAS.471.3152P}, but we find a smaller amplitude, probably as a result of less efficient grain growth in galaxies with low stellar mass (see discussion in Section \ref{sec:DSM}). The differences with \citet{2019MNRAS.485.1727H} are likely due to a combination of different simulation volumes/resolution\footnote{The DMFs computed by  \citet{2019MNRAS.485.1727H} are based on simulations with 50 $h^{-1}$~Mpc$^3$ and 512$^3$ particles, hence they simulate a larger volume but have less resolution compared to our RefRun (their dark matter and gas particle masses are $6.98\times10^7 h^{-1} M_\odot$ and $1.28 \times 10^7 h^{-1} M_\odot$, respectively).} and sub-grid prescriptions for grain growth. In fact, the adopted rate 
has a similar functional form, but it is implemented in a different ISM phase. In \texttt{dustyGadget}, grain growth occurs only in star-forming particles, when their number density exceeds a threshold of $132 h^{-2}$~cm$^{-3}$. In addition, its timescale is modulated with the gas metallicity and the number density of star forming particles, having a certain $x_c$; this implies that only when a galaxy has reached a considerable cold phase in star forming regions, the dust can efficiently increase in mass. In \citet{aoyama2018} grain growth occurs in dense gas particles, classified as particles with 
$n > 10$~cm$^{-3}$, with a timescale that is modulated only by the gas metallicity and which corresponds to $\tau_{\rm gg,0}(n_c,T_c) = 1.2$ Myr\footnote{This value is obtained by rescaling $n_c$ in $\tau_{\rm gg}$ to 1000~cm$^{-3}$ instead of 100, as in the original paper.}.
The selection of these regions likely favours the evolution of a larger sample of dusty dwarfs, and disfavours objects with dust masses higher than $M_{\rm d} \sim 10^{7.5}$~$M_{\odot}$. 

\begin{figure*}
\centering
\includegraphics[angle=-90,scale=0.75, trim={0.3cm 0cm 0.40cm 0.2cm},clip]{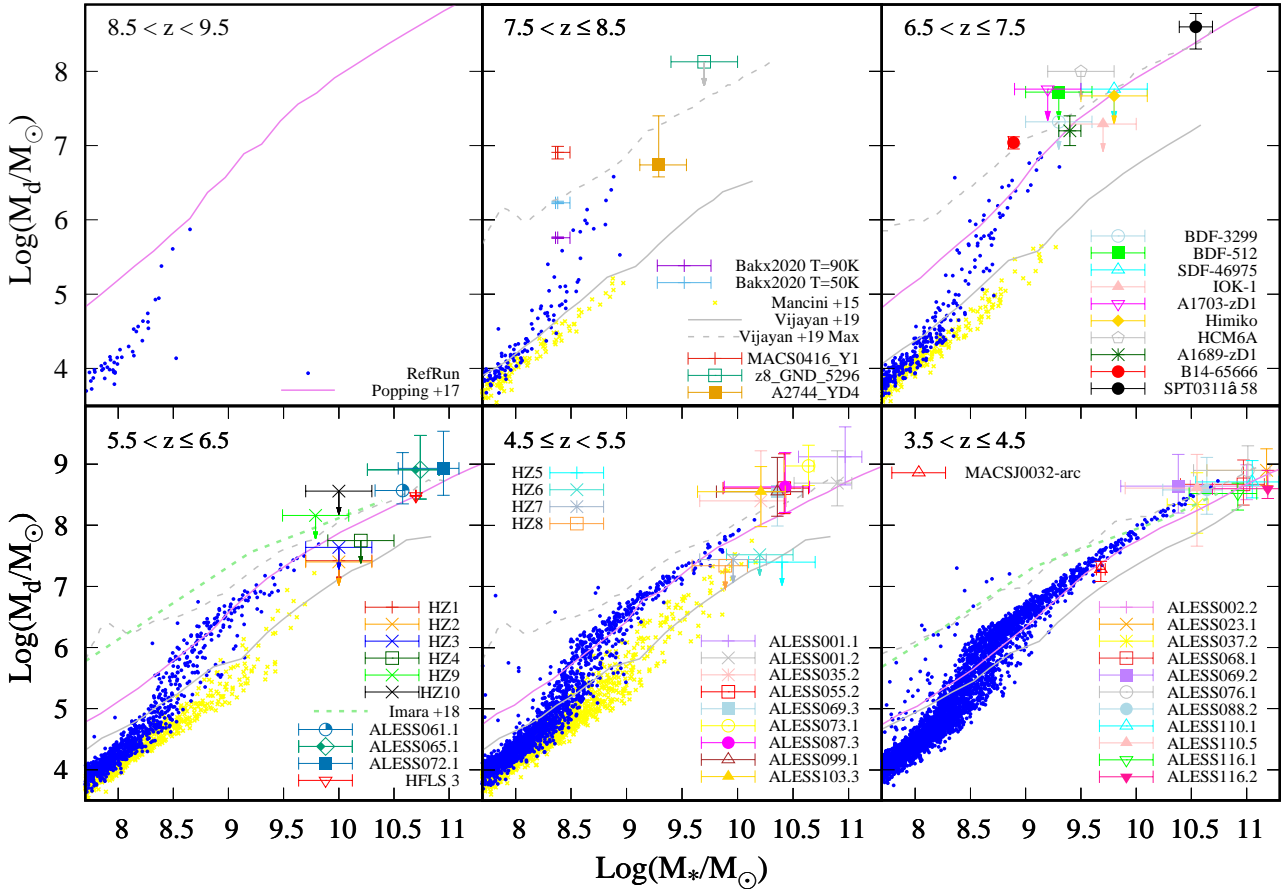}
\vspace{0.01truecm}
\caption{Logarithm of the dust mass $M_{\rm d}$~[$M_{\odot}$] as a function of the stellar mass $M_{\star}$~[$M_{\odot}$] for the galaxy sample with Log$(M_{\star}/M_{\odot}) > 7.5$. The redshift evolution of the galaxies is shown from $z\approx 9$ (top left panel) down to $z\approx 4$ (bottom right). In all panels, blue points are galaxies found in the reference run (RefRun) while the same objects predicted by the semi-numerical model of \citet{2015MNRAS.451L..70M} are shown as yellow points. When available, we also show the results of independent studies: 
the average trends computed by \citet{2017MNRAS.471.3152P} are shown as solid violet lines, gray solid and dashed lines are the fiducial/max models by \citet{2019arXiv190402196V} and dashed green lines are the predictions by \citet{2018ApJ...854...36I}. The observed dust and stellar masses can be found in Table \ref{tab:ObsTable}.}
\label{fig:Fig3MdMstar}
\end{figure*}
\begin{figure*}
\centering
\includegraphics[angle=-90,scale=0.75, trim={0.3cm 0cm 0.40cm 0.2cm},clip]{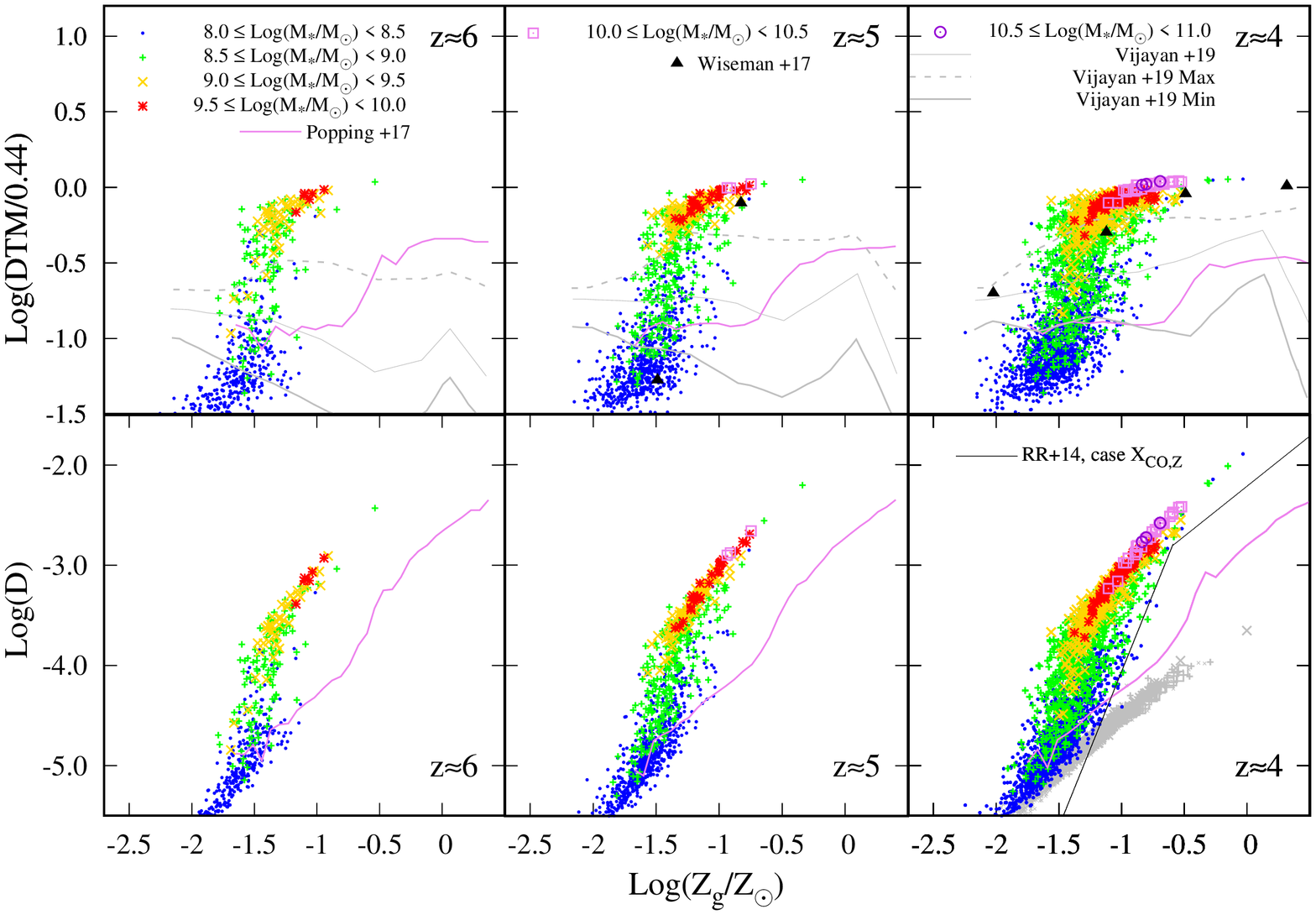}
\vspace{0.01truecm}
\caption{Redshift evolution of Log$({\rm DTM/0.44})$ (top panels) Log${\cal D}$ (bottom panels) as a function of Log$(Z_{\rm g}/Z_{\odot})$, where $Z_{\rm g}$ is the gas phase metallicity in solar units (see text for more details). The points refer to the RefRun and show galaxies in the stellar mass range $8 \leq {\rm Log}(M_{\star}/M_{\odot}) \leq 11$, grouped in bins of one order of magnitude and shown with different symbols and colors (see the legend in the top left panel). For comparison, we also show the predictions of \citet{2017MNRAS.471.3152P} (solid violet line), \citet{2019arXiv190402196V} (solid gray line for the fiducial case and dashed gray line for the maximum dust production efficiency case); the triangles show data from \citet{2017A&A...599A..24W}. Finally, in the bottom-right panel ($z\approx4$) we also show  Log${\cal D}$ derived from the sample of the "ProdOnly" simulation, with identical symbols but gray colors and the fit in \citet{remyruyer2014} (solid black line for the broken power law case, X$_{\rm CO,Z}$), valid for a galactic sample at $z=0$.}
\label{fig:Fig4DGRatioZ}
\end{figure*}

\subsubsection{Dust-to-stellar mass}
\label{sec:DSM}

Here we explore how dust and stellar mass correlate in the galactic sample of the RefRun as function of redshift, i.e. the relation $M_{\rm d} (M_{\star}(z))$. Figure \ref{fig:Fig3MdMstar} shows this evolution in different panels from $z=9$ (top left) to $z=4$ (bottom right panel), by selecting objects with $M_{\star} > 10^{7.5}$~ $M_{\odot}$  (solid blue points,  because smaller galaxies are poorly resolved given the currently adopted mass resolution. The data reported in Table 2 and in the figure show that current targets of ALMA observations have masses $M_{\star} \geq 10^{8.5}$~ $M_{\odot}$. Each $z$ in this figure is the middle value of a redshift interval $\Delta z =1$, in which the observed sample of  Table \ref{tab:ObsTable} is also grouped, for a better comparison\footnote{Note that when more than one value of $M_{\rm d}$ for a single object is present in the table, we only show $M_{\rm d}(T_{\rm d} \approx 35 K)$.}. Yellow crosses refer to the semi-numerical model by \citet{2015MNRAS.451L..70M} and \citet{2016MNRAS.462.3130M}, where galaxies are identified using the \texttt{AMIGA} halo finder and have the same stellar masses and metallicity as in the present \texttt{dustyGadget} simulation. 

At all redshifts, dust enrichment at the lowest mass end is dominated by stellar sources, consistent with the findings of \citet{2015MNRAS.451L..70M, 2016MNRAS.462.3130M}. As expected, in this regime the dust mass grows linearly with the stellar mass, as a result of the equilibrium between dust formation by SNe and AGB stars and dust destruction by SN shocks. As larger mass galaxies assemble, their cold ISM phase is more favourable to grain growth and the dust mass increases rapidly with stellar mass, reaching a saturation when accretion is limited by dust destruction and by 
the gas phase metallicity. These features are particularly clear at $z \leq 6$ and are very consistent with the masses indicated by the ALESS sample and with what is observed locally in samples of galaxies spanning a sufficiently large range of metallicity \citep{2013EP&S...65..213A, remyruyer2014, 2014MNRAS.445.3039D, 2014A&A...562A..76Z, 2018MNRAS.473.4538G}. Also note the excellent agreement with MACSJ0032-arc, a smaller object with $\rm SFR \approx 50$~$M_{\odot}$~yr$^{-1}$ observed at $z \approx 3.6$.

At $z > 6$, most of the simulated galaxies have low $M_{\star}$ but there is a clear trend of increasing $M_{\rm d}$ in the most massive galaxies at each redshift, above the simple extrapolation of the linear regime. This is very interesting as it shows that, provided the right conditions are met, grain growth can operate even in relatively chemically unevolved galaxies. At these high redshifts, the major driver is likely to be the mass and density of the cold gas phase: indeed, the results of  \texttt{dustyGadget} always lie above the predictions of \citet{2015MNRAS.451L..70M} where no density modulation of the grain growth timescale was considered. 

The dust masses in our largest simulated galaxies are consistent with the observationally inferred values at all redshifts, although at $z > 5$ the statistics at the high mass end is poor, due to a combination of two effects:
({\it i}) the relatively small simulated volume and ({\it ii}) a lack in mass resolution, which result in an artificial underestimation of star formation at high redshift, as discussed by \citet{2019MNRAS.484.1852A}.
Hence, a direct comparison with the data is not possible in some redshift bins. An example is provided by the $7.5 \leq z < 8.5$ bin (top middle panel). None of our galaxies has reached the right stellar mass to allow a direct comparison with A2744\_YD4, although the trend of the most massive objects in our sample seems to be in line with its $M_{\rm d}$. 
In addition, we point out that the observationally inferred dust masses are significantly affected by the adopted dust temperature and emissivity properties. As an example, in the top-middle panel we report
the dust mass estimated for the $z = 8.3$ galaxy MACS0416\_Y1 by \citet{2019ApJ...874...27T} and the recently revised values reported by \citet{Bakx2020} who provide a dust mass range that is remarkably close
to our predicted values at comparable redshifts and stellar masses.

Interestingly, the simulated galaxies in the redshift range $6.5 \leq z < 7.5$ easily match the upper limits on $M_{\rm d}$ inferred for normal star forming galaxies\footnote{In this redshift interval none of the normal star forming galaxies has a direct detection of the IR continuum from which the dust mass can be derived; the upper limits are then derived from \citet{2015A&A...574A..19S} by assuming $T_{\rm d}=35$~K.} but - at the same time - they are also consistent with the dust mass inferred for A1689-zD1. Hence, our environment-dependent dust enrichment naturally accounts for the variety of objects that have been observed at these high redshifts, some of which have a direct detection of their rest-frame IR continuum while for others we can rely only on upper limits, even in very deep exposures. This is particularly encouraging, given that in previous studies the grain growth timescale required to match this data had to be artificially reduced by one order of magnitude ($\tau_{\rm gg,0} = 0.2$~Myr) and it was argued that the more efficient grain growth was due to a larger ISM density \citep{2015MNRAS.451L..70M}.

Our data is also compared with average trends in the redshift interval $4<z<9$ predicted by the semi-analytic models of  \citet{2017MNRAS.471.3152P} (solid violet line), \citet{2019arXiv190402196V} (solid gray line for the fiducial case and dashed gray line for the maximum dust production efficiency case) and \citet{2018ApJ...854...36I} (dashed green line). In general, the agreement is very good at the high mass end, where our simulated galaxies have already entered the regime where grain growth starts to be efficient. Conversely, at all $z$ the semi-analytical models predict a larger dust mass 
at the low-mass end compared to \texttt{dustyGadget}, probably due to differences in the adopted stellar dust yields or grain condensation efficiency. 

\begin{figure*}
\centering
\includegraphics[angle=0,scale=0.45, trim={0.3cm 0cm 0.40cm 0.1cm},clip]{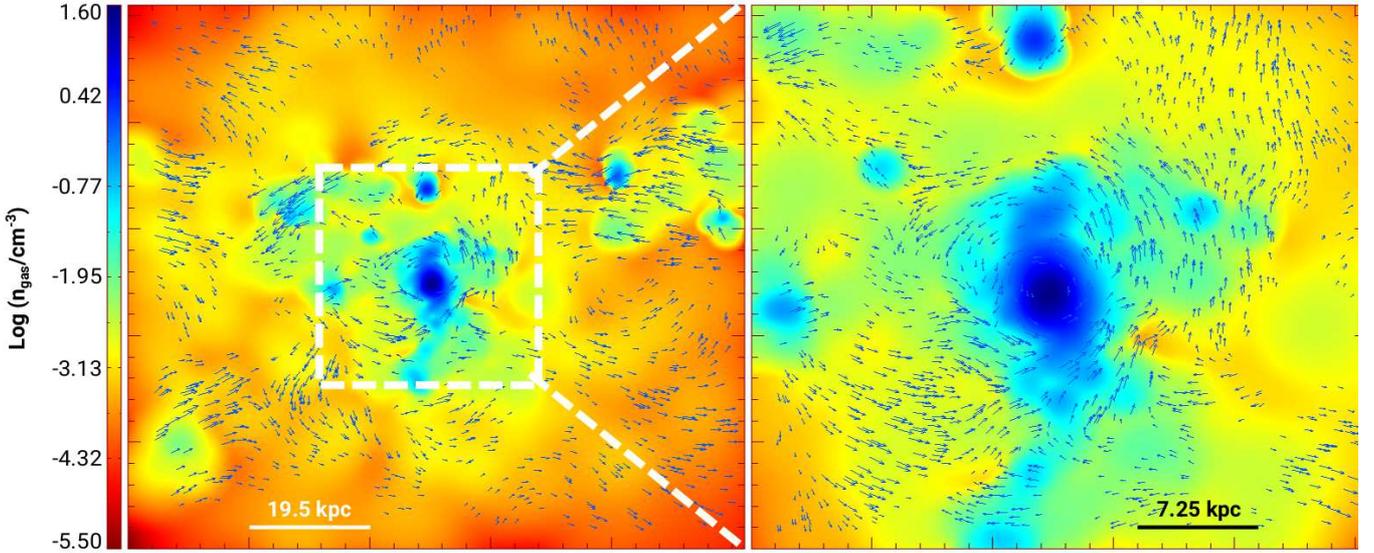}
\vspace{-1.00truecm}
\caption{Gas distribution in the selected  DM0 halo at $z\approx4$. The gas number density $n_{\rm gas}$~[cm$^{-3}$] is mapped on a grid of 256$^3$ cells covering a cosmic volume of $L_b \approx 100$~kpc, physical, centered on the central galaxy (G0). The {\bf Left panel} shows a slice cut of the volume containing the center of G0, while the {\bf Right panel} zooms into a central sub-region of size $\approx 37$~kpc. In both panels the logarithmic value of $n_{\rm gas}$ is shown as color palette from red (low density gas) to dark blue, while the velocity field of the gas is shown by blue arrows of different length spanning the velocity module interval [50-300]~km s$^{-1}$.}
\label{fig:Fig5nGasG0}
\end{figure*}

\subsubsection{Dust-to-metal and dust-to-gas ratios}
\label{sec:Ratios}

In this section we first study the average dust-to-metal ratio (DTM) of the galaxies found in the RefRun. This is defined as ${\rm DTM} \equiv M_{\rm d}/M_{\rm z,g}$, where $M_{\rm d}$ is the total mass of dust in each galaxy and $M_{\rm z,g}$ is the total mass of metals in gas phase. Second, we focus our attention on the analogously defined, dust-to-gas ratio ${\cal D}$. Both quantities are discussed as function of the gas phase metallicity $Z_g$, defined as total mass of gas-phase metals over total mass of gas ($Z_{\rm g} = M_{\rm z,g}/M_{\rm g}$)\footnote{To show this quantity in solar units we adopt the solar metallicity value $Z_{\odot} = 0.014$ \citep{2009ARA&A..47..481A}.}. Figure \ref{fig:Fig4DGRatioZ} shows the redshift evolution of  Log$(\rm DTM/0.44)$\footnote{For a better comparison with data from other models, the DTM ratio is normalized with respect to its value in the Milky Way, $\rm DTM_{\rm MW}=0.44$, adopted in \citet{2017MNRAS.471.3152P}.} in the top panels and Log${\cal D}$ in the bottom panels. To facilitate a direct comparison with Figure \ref{fig:Fig3MdMstar}, the data points refer to a stellar mass range $8.0 \leq {\rm Log}(M_{\star}/M_{\odot}) \leq 11.0$. Galaxies are also grouped in bins of stellar mass and shown with different symbols and colors (see the legend in the top panels). In the bottom-right panel ($z\approx4$) we also show  Log${\cal D}$ of the "ProdOnly" simulation, with identical symbols but in gray-to-black color gradient. Data from \citet{2019arXiv190402196V} (ref/Max case) are shown as coloured lines with the usual graphic style. Values based on observations available from \citet{2017A&A...599A..24W} at $z=4.5$ are finally shown as black triangles.

Our points highlight a clear metallicity evolution in redshift from $z=6$ to $z=4$ following the assembly of larger mass galaxies (see the progression in the coloured points from blue to red and magenta). The DTM and ${\cal D}$ show a similar behaviour, quickly rising with $Z_{\rm g}$ at all $z$. The interpretation of these trends follows from the discussion of Figure \ref{fig:Fig3MdMstar} and the mass-metallicity relation. Indeed, at low $Z_{\rm g}$ (low stellar masses), the dust content in galaxies is set by the balance between dust production by stellar sources and dust destruction. As $Z_{\rm g}$ increases, grain growth starts to be efficient, leading to a rapid increase in both DTM and ${\cal D}$. Finally, above Log($Z_{\rm g}/Z_{\odot})\approx -1.5$, the dust content reaches an equilibrium controlled by grain growth and dust destruction. As a consequence, ${\cal D}$ becomes linearly proportional to the metallicity and DTM reaches a saturation, that persists across redshifts up to the highest metallicity  (Log$(Z_g/Z_{\odot})\approx -0.5$, i.e. $\approx 0.3 Z_{\odot}$). Interestingly, the DTM of the simulated galaxies appear to be in very good agreeement with the few available observations at $z = 5$ and $z = 4$ \citep{2017A&A...599A..24W}. 
First, the two data points at $z=5$ (top middle panel) confirm that a significant increase in DTM can be observed in galaxies with a moderate difference in $Z_{\rm g}$. Second, at $z\approx 4$ where more data are available (top right panel), observations are very well reproduced by \texttt{dustyGadget}.

A large discrepancy across models appears in the top panels. While the DTM and ${\cal D}$ of \citet{2017MNRAS.471.3152P} follow the same general trends predicted by \texttt{dustyGadget}, a rapid increase in their values occurs at high gas metallicity. At low metallicity (low stellar masses) their predicted ${\cal D}$ are very consistent with our results. Since Figure \ref{fig:Fig3MdMstar} shows larger dust masses for galaxies with Log$(M_\star/M_\odot) < 8.5$, we conclude that some of the differences may be due to a gas content in objects predicted by \citet{2017MNRAS.471.3152P} larger than the one in our simulated systems.

\subsection{The dusty environment of a massive halo at $z\approx4$}
\label{sec:dustyMMHalo}
Thanks to its full hydrodynamical approach \texttt{dustyGadget} is also capable to provide accurate information on the distribution of dust within and around galaxies.This information can be compared with resolved observations and it is of great importance as it affects both luminosity and colours of observed galaxies.  

With this aim in mind, here we qualitatively investigate the environment of one massive halo, DM0, with a total dark matter mass of $M_{\rm DM} \approx 1.2 \times 10^{12}$~$M_{\odot}$ and a virial radius of $R_{\rm vir} \approx 250$~kpc 
at $z \approx 4$, for which our present simulation is able to achieve an adequate spatial resolution\footnote{The halo is composed of 14249 DM particles, 12834 gas particles and 8580 stellar particles. It was selected among the most massive ones found at $z \approx 4$ not having a major merger event.}.
The halo contains a central galaxy (G0) with gas mass $M_{\rm gas} = 1.0 \times 10^{11}$~$M_{\odot}$, stellar mass $M_{\star} = 2.4 \times 10^{10}$~$M_{\odot}$, and dust mass $M_{\rm d} = 1.62 \times 10^8 M_\odot$, producing stars at a rate of SFR $\approx 32$~$M_{\odot}$ yr$^{-1}$. 
These properties are intermediate between those observed for MACSJ0032-arc and for the lowest star forming galaxies of the ALESS sample (e.g. ALESS088.2 or ALESS110.5). While a closer comparison with single objects is deferred to a future simulation with higher mass resolution, here we aim at describing a typical dusty environment found at the high-end of our $M_{\rm d} - M_{\star}$ relation at $z = 4$. 

\begin{figure*}
\centering
    \subfloat[]{
    { \includegraphics[angle=0,width=0.45\textwidth,clip]{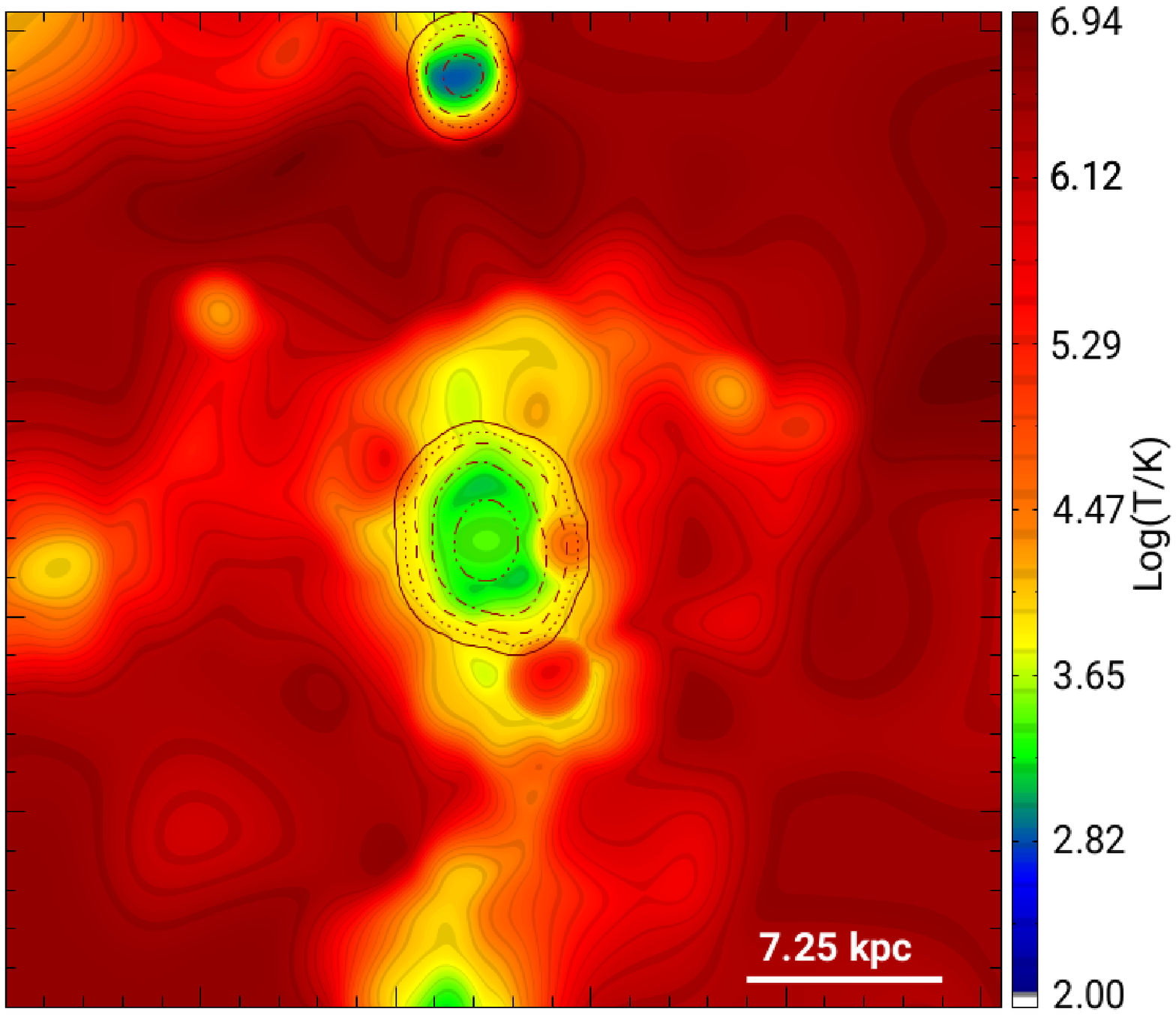}
    \label{fig:Fig6T}}
    }%
    \qquad
    \subfloat[]{
    {\includegraphics[angle=0,width=0.45\textwidth,clip]{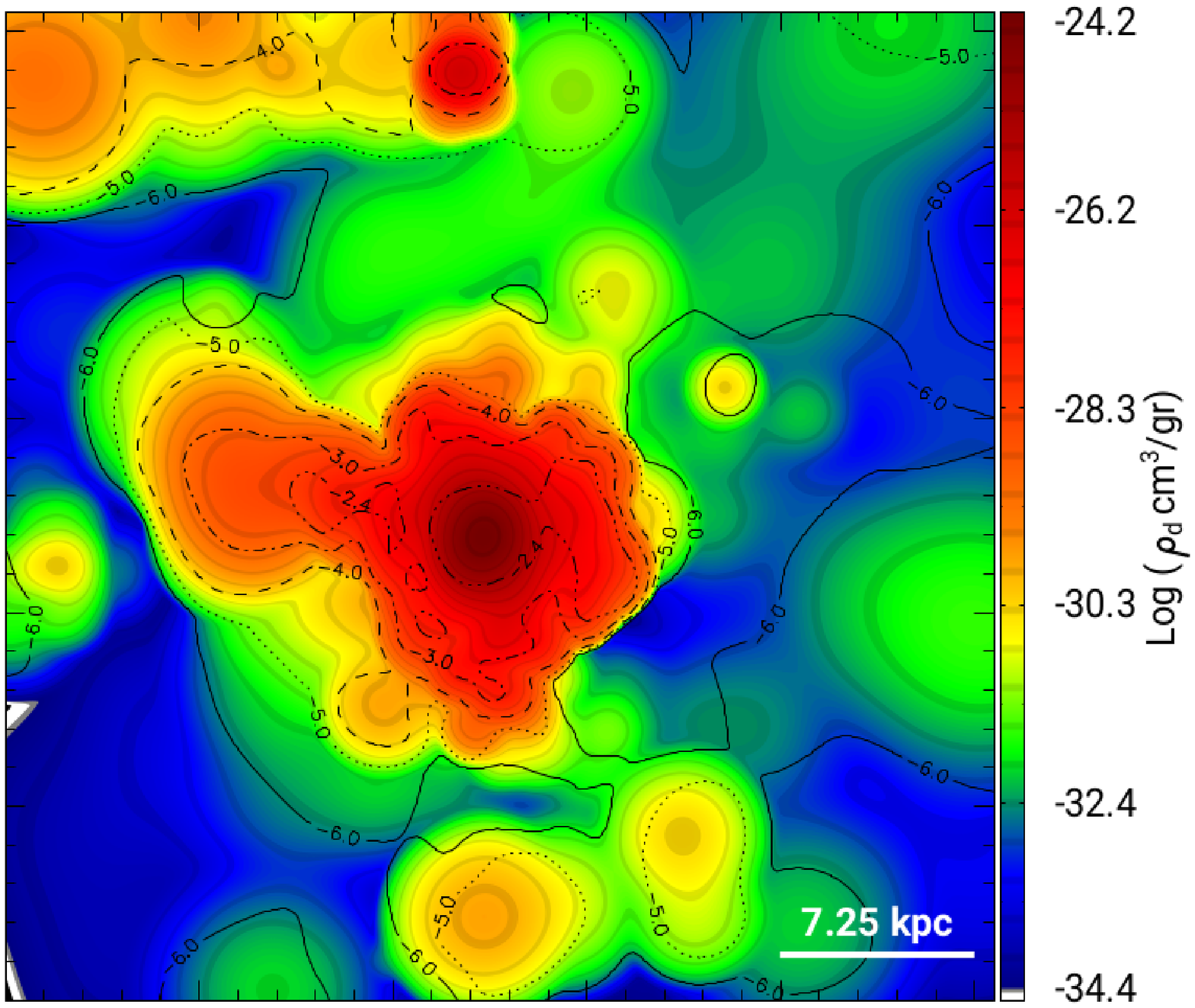}
    \label{fig:Fig7DG}}
    }%
\caption{Gas temperature and dust density distributions in the same Cartesian grid and slice cut of the right panel in Figure \ref{fig:Fig5nGasG0} in a box of $\approx37$~kpc. {\bf Left Panel (a)} Logarithm of the gas temperature $\rm Log(T_{\rm gas}/K)$ shown as a color palette ranging from blue ($T_{\rm gas}\approx 10^2$~K) to dark red ($T_{\rm gas} > 10^6$~K). Values of the cold gas number density (Log$(n_c)$) are over-plotted as black lines with different styles: Log$(n_c) = -4$ (solid),   -2 (dotted),  -1 (dashed), 0 (dashed-dotted), 1 (dashed-triple-dotted). {\bf Right Panel (b)} Logarithm of the dust density  Log$(\rho_{\rm d}/{\rm gr\,cm^{-3}})$ shown as a color palette ranging from dark blue (negligible content) to dark red  (Log($\rho_{\rm d}/{\rm gr\,cm^{-3}}) = -24$). Values of the logarithm of the dust-to-gas ratio (Log$({\cal D})$ are over-plotted as black lines with different styles: Log$({\cal D})$ = -6 (solid), -5 (dotted),  -4 (dashed),  -3 (dashed-dotted), -2.4 (dashed-triple-dotted).}
\end{figure*}

DM0 also contains 12 luminous satellites ($6 \leq {\rm Log}(M_{\star}/M_{\odot}) \leq 8$) orbiting the central galaxy at a physical distance in the range $9<d/{\rm kpc}<50$ and two dark satellites polluted by dust.

The dynamics of the gas present in the environment of G0 is shown in the left panel of Figure \ref{fig:Fig5nGasG0}. In particular, we visualize its number density ($n_{\rm gas}$) on a slice cut of a box centered on G0, and with a side length $L_b\approx 100$~kpc. To find the spatial distribution of the gas, the SPH particle distribution has been projected onto a Cartesian grid of 256 cells/side, corresponding to a spatial resolution of 0.39~kpc; Log$(n_{\rm gas}/{\rm cm^{-3}})$ is then shown as color palette from red (Log$(n_{\rm gas}/{\rm cm^{-3}}) \approx -5.5 $, typical of inter-galactic environments) to blue (Log$(n_{\rm gas}/{\rm cm^{-3}}) \approx 1.6$, a galactic ISM density). 

Despite the selection effect due to the geometric cut, some of the satellites are visible as blue, dense systems, often connected to the central object by cyan/green circum-galactic gas streams with $0.01$~cm$^{-3}< n_{\rm gas} \leq 0.2$~cm$^{-3}$. Over a distance of $d \approx 50$~kpc, and up to the box boundaries, the gas becomes more diffuse and its number density rapidly drops toward values closer the ones of the large scale, i.e. n$_{\rm gas} < 10^{-4}$~cm$^{-3}$. 

Finally, the velocity field of the gas, relative to the one of G0, is shown by blue vectors with length proportional to their module\footnote{As for the number density, the velocity field has been projected on the same grid by mass weighting the velocity of each particle belonging to a grid cell.}. For the sake of clarity, the vectors are shown in representative sub-regions and only if their absolute value is $|\vec{v}| > 40$~km/s. 
Large scale motions are present almost everywhere in the box, with a complex pattern connecting the central galaxy with its surroundings. 

To better resolve the dusty environment of G0, we further zoom-in a smaller, central volume of side length $L_{b,c} \approx 37$~kpc covered by a new grid of 256 cells/side on which the original gas particles are projected. The resulting spatial resolution of the central region (indicated as a dashed square in the left panel) is $d_c\approx 0.144$~kpc and its gas distribution is shown in the right panel of Figure \ref{fig:Fig5nGasG0}.

The central, disk-like galaxy, G0, is better resolved on the scales presented in the right panel and it shows a gas distribution which drops to $n_{\rm gas}\approx 10^{-2}$~cm$^{-3}$ within $r_{\rm gas}\approx 12$~kpc (i.e. the green area). The central, dense area is quite compact and corresponds to the bulk of the star forming ISM of the galaxy. It should be noted that while our mass resolution is not adequate to resolve the ISM clouds of the central region, it is sufficient to provide a first indication on how the cosmic dust spatially correlates with the distribution of the gas.

Figure \ref{fig:Fig6T} shows the mass weighted gas temperature on the same scales\footnote{Note that this is the \texttt{dustyGadget} temperature of the SPH particles projected on the grid by weighting for their mass. To avoid  quantization effects due to the SPH particle mass resolution on the gas properties, the grid has been selected to ensure that many gas particles contribute to the mass weighted value in each cell. The yellow regions shown in the figure, for example, are extremely hot/dense cells to which many particles belong to. Finally note that a kernel-smoothing is also applied when the data is plotted. Radiative effects are not accounted for at this scale with sufficient detail as our SPH scheme only accounts for an extragalactic UV background field, tuned on the large scale. This is not appropriate/significant at the galactic scale where the contribution of each star forming regions should be accounted for. In a future work we will investigate this point by adopting a better mass and spatial resolution and performing radiative transfer simulations accounting for dust as in \citet{2019MNRAS.482..321G}.}. 

The inner region of the galaxy ($d_{\rm G0} \approx 2.5$~kpc) is dominated by warm gas with $T_{\rm gas}\approx 10^3$~K (green-yellow areas\footnote{Also compare with iso-contours of the cold gas mass, see figure caption for more details.}). $T_{\rm gas}$ progressively increases up to $T_{\rm gas}\approx 10^4$~K in a intermediate region corresponding to the galaxy surroundings (light-red areas here, cyan-green patterns in Figure \ref{fig:Fig5nGasG0}), while the outskirts of the halo are dominated by a low density, shocked gas (dark red here, green-yellow regions in Figure \ref{fig:Fig5nGasG0}) with temperatures as high as $T_{\rm gas}\approx 3 \times 10^6$~K and densities lower than $n_{\rm gas} < 10^{-3}$~cm$^{-3}$. 

It is interesting to compare the above maps with the dust distribution predicted in the same region. This is shown in Figure \ref{fig:Fig7DG}, where the logarithm of the dust density ($\rho_d$ in ${\rm gr \,cm}^{-3}$) is presented with a colour palette from dark blue (negligible content) to dark red 
($\rm Log (rho_d/{\rm gr \, cm^{-3}}  = -24$). Iso-contours of Log \,${\cal D}$ are superimposed on this picture as black lines of different line style, ranging from  Log \,${\cal D} = -6.0$ (solid lines) to Log\,${\cal D} = -2.4$ (triple-dotted-dashed lines).

By comparing Figure \ref{fig:Fig6T} and \ref{fig:Fig7DG}, it appears that dust spatially correlates only with the cold gas, where it is injected
in the ISM by stars and where grain growth can occur. As a result, the spatially resolved dust-to-gas ratio is ${\cal D} > 3 \times 10^{-2}$, slightly higher than the average value measured in the Milky Way; note that the average value in the halo is ${\cal D} > 1.62 \times 10^{-3}$. The spatial distribution of the cold phase (iso-contour lines in Figure \ref{fig:Fig6T}) is
more regular and centrally concentrated than that of dust grains (iso-contours in Figure \ref{fig:Fig7DG}), indicating that winds are at place in the inner, star forming regions and cause an inhomogeneous enrichment of metals and dust. The dust density rapidly decreases in the outskirts of the galaxy creating a very diffuse and irregular pattern (yellow and green regions). This shows that dust is present everywhere in the halo and often pollutes filaments connecting the central galaxy with its surrounding satellites. Its density is, on the other hand, orders of magnitude lower than in G0 ($\rm Log (rho_d/gr \, cm^{-3}  = -30$) and the grains are extremely diluted in the inter-galactic gas, with $D < 10^{-3}$. 

Dust grains moving into regions with very high temperature can suffer efficient sputtering decreasing their mass. Indeed, their spatial distribution appears to be anti-correlated with the location of gas shocked regions (see green-blue areas in Figure \ref{fig:Fig7DG} at $T_{\rm gas} > 10^5$~K).

\section{Conclusions}
\label{sec:dustConclusions}

In this paper we investigated the formation and chemical evolution of a sample of dusty galaxies found in a cosmological volume of comoving box size of $30h^{-1}$~cMpc. The sample is simulated with \texttt{dustyGadget}, an extension of the SPH \texttt{Gadget} code, capable of accounting for dust production and evolution. Dust production is modeled using mass and metallicity-dependent stellar yields of AGB stars and SN accounting for the effects of the reverse shock. Dust grains are evolved in the galactic ISM, consistently with its hot and cold phases, through processes of destruction and grain growth by accretion of metals from the gas phase.  

The simulated galaxy properties are compared with both independent theoretical predictions and with observed samples of galaxies at $z\geq4$. From this analysis, we find that:

\begin{itemize}  

\item The evolution of the cosmic dust density parameter is driven by stellar dust production at $z \geq 10$. Accurate modeling of stellar dust yields, including those produced by metal-free and very metal-poor stars, is therefore very important in order to assess the mass, composition and spatial distribution of interstellar dust in $z \approx 10$ systems. \texttt{dustyGadget} is particularly suited for this purpose, as it follows metal and dust enrichment on stellar characteristic lifetimes, starting from a metallicity-dependent stellar initial mass function, and describing their chemical (AGB, SN, PISN) and mechanical (SN, PISN) feedback across a wide range
of stellar masses. At $z < 10$, grain growth in the cold neutral phase of metal-enriched galaxies starts to be efficient, driving the evolution of $\Omega_{\rm d}$ 
towards a value of $\approx 10^{-6}$ at $z = 4$, in good agreement with independent studies \citep{aoyama2018} and about an  order of magnitude higher than predicted by stellar sources only.

\item In agreement with previous studies \citep{2015MNRAS.451L..70M, 2016MNRAS.462.3130M}, we find that, at any redshift, interstellar dust in low-mass galaxies, with Log$(M_\star/M_\odot) < 8.4$, is largely produced by AGB and SNe, and the dust mass grows linearly with stellar mass. Across the mass range $8.5 < M_\star/M_\odot < 9.5$ the dust mass rapidly increases, with a wide dispersion, due to grain growth. We find that when the grain growth timescale is computed consistently with the properties of the cold gas phase, the resulting dust mass at the high mass end of the simulated galaxy distribution is in good agreement with the values inferred from either direct detections or deep upper limits of the rest-frame IR continuum in galaxies in the redshift range $ 4 \leq z < 8$. This confirms previous indications of a density dependence of the grain growth timescale, found in models \citep{2015MNRAS.451L..70M, 2017MNRAS.471.3152P} and observations \citep{schneider2016, romanduval2017}. 

\item Although independent models show a fairly good agreement in predicting the evolution of the cosmic dust density parameter in the redshift range $4 < z < 6$, they differ in predicting the dust mass function at $z = 4$ and 5. This may be ascribed to differences in galaxy samples due to variations in simulation boxes/resolution, in intrinsic galaxy properties (particularly the mass fraction in cold gas), and in the sub-grid implementation of grain growth. These differences are also reflected in the metallicity-dependence of the dust-to-metal and dust-to-gas ratios, which greatly vary across models. It is encouraging that our simulated galaxies at $z = 5$ and 4 appear to have dust-to-metal ratios consistent with the few available observations \citep{2017A&A...599A..24W}.

\item Despite the limited resolution achieved in our simulation, a qualitative investigation of the properties of the most massive halo at $z\approx 4$ shows a complex gas distribution, connecting the central most massive disk-like galaxy to its satellites through low-density filaments. Dust grains appear to spatially correlate with the cold gas. In the innermost 2.5~kpc around the central massive galaxy, the dust-to-gas ratio is ${\cal D} > 0.03$, larger than the average value found in the Milky Way. 
The grains clearly escape the galaxy through galactic winds, reaching physical distances of $30$~kpc from the central object. An intricate pattern connecting the central galaxy with its dusty, inefficiently star-forming satellites is clearly shown by our maps. 
\end{itemize}

Although more detailed comparison with individually detected objects are deferred to future investigations, our results suggest that - provided the right conditions are met - dust enrichment can proceed rapidly at high redshift, aided by star formation and grain growth in the gas-rich regions of the first galaxies.

\section*{Acknowledgments}
We thank the Referee, S. Aoyama, for his careful reading of the manuscript and constructive comments. We also thank G. Popping for the support in the model comparison, M. Palla, F. Matteucci and C. Peroux for useful discussions. The authors also thank V.Springel and K.Dolag for supporting the development of 
\texttt{dustyGadget} from the good, old \texttt{Gadget2} and allowing access to the 
code for a future \texttt{Gadget3} porting. The research leading to these results has received funding from the European Research Council under the European Union's Seventh Framework Programme (FP/2007-2013) / ERC Grant Agreement n. 306476. 
LG acknowledges support from the Amaldi Research Center funded by the MIUR program "Dipartimento di
Eccellenza" (CUP:B81I18001170001). LG, MG, LH acknowledge funding from the INAF PRIN-SKA 2017 program 1.05.01.88.04. UM is supported by the German Research Fundation (DFG), project n. 390015701 and the HPC-Europa3 Transnational Access Programme, project n. HPC17ERW30. 

\bibliographystyle{mn2e}
\bibliography{dustyGadget}

\begin{appendix}

\setcounter{table}{0}
\renewcommand{\thetable}{A\arabic{table}}

\begin{table*}  
\begin{tabular} {|l|l|l|l|l|l|l} \hline

                          & Popping+17 & Mancini+ 15 & Gioannini+ 17    & Aoyama+18  & McKinnon+ 17& \texttt{dustyGadget}\\
\hline
Cosmology                  & WMAP5       & WMAP7       & Planck 2016     & Planck 2016 & Planck 2014 & WMAP7           \\  
L$_b$ [cMpc~$h^{-1}$]      &   -         & 30          &     -           & 50          & 25          & 30              \\
M$_{\rm DM}$ [M$_{\odot} h^{-1}$] & -    & $6\times10^7$ & - & $6.89\times10^7$ & $8.22\times10^6$ & $6\times10^7$     \\
M$_{\rm g}$ [M$_{\odot} h^{-1}$]  & -    & $9\times10^6$ & - & $1.28\times10^7$ & $1.53\times10^6$ & $9\times10^6$     \\
ISM Phases                 &diff/H$_2$ $^b$  & x$_c=0.5$    & x$_c*$   & f$_{\rm dense}=0.1$ $^d$& -   & Hot/cold $^a$\\
Dust sources               &SN Ia-II/AGB & SN II/AGB  & SN II/AGB       & SN Ia-II/AGB & SN Ia-II/AGB & PISN/SN II/AGB \\
RS in yields               &Y (avg) $^i$ &Y (tbl.) $^i$& Y (avg) $^h$   & N           & N           & Y (tbl) $^i$     \\
Astration                  &Y            &Y           & Y               & Y           & Y           & Y                \\
$\tau_{\rm gg,0}$~[Myr]    &1.5          &2.0         & 2.0             & 1.2         & 4           & 2.0              \\
$f_{\rm SN}$               &0.36         &0.15        & 1               & -           & 1           & 0.15             \\
$\epsilon_{\rm d} $        &*            &0.48        & 0.1             & 0.1         & 0.3         & 0.48             \\
M$_{\rm s}$~[$M_{\odot}$]  &600-980*     &2040        & 1360            & ?           & 7412        & 2040             \\
GS distribution            &N            &N           & N               & Y           & N           & N                \\                                    
\hline
\hline
\end{tabular}
\caption{Properties of semi-analytical, semi-numerical, and numerical models where dust evolution has been implemented. The parameters assumed for the various dust processes are extracted from the original paper models and references therein (see Table \ref{tab:models}). References in this table: $a$: \citet{2003MNRAS.339..289S,2010MNRAS.407.1003M}, $b$: \citet{2015MNRAS.453.4337S}, $d$: \citet{2017MNRAS.466..105A}, $i$:  \citet{2007MNRAS.378..973B}, $h$: \citet{2011arXiv1107.4541P}.}
\label{tab:modelValues}
\end{table*} 

\section{Multiphase ISM model}
\label{sec:AppAISMModel}

\citet{2003MNRAS.339..289S} implemented a sub-resolution model which uses spatially averaged physical properties to describe a two-phase ISM contained in each SPH gas particle. In their picture, the two phases (hot and cold gas)\footnote{In this paper variables with a subscript $_{\rm c}$ refer to all the variables of the cold phase, while $_{\rm h}$ indicates the hot one.} survive in pressure equilibrium as prescribed by the equations introduced in \citet{1977ApJ...218..148M}. 

Volume-averaged quantities, as for example the gas density ($\rho_g$) and temperature ($T_g$), can be simply  related to the average  values of each phase, i.e.  $\rho_g = \rho_{\rm h} +  \rho_{\rm c}$. Similarly, the average thermal energy of the gas per unit volume can be written as:   $\epsilon_g =  \rho_{\rm h} u_{\rm h}  +  \rho_{\rm c} u_{\rm c}$, where $u_{\rm h}$ and $u_{\rm c}$ are the energy per unit mass of the hot and cold components, respectively.

The cold phase represents condensed clouds where stars form and it is assumed at constant temperature $T_c \sim 10^3$~K (i.e. constant $u_{\rm c}$), while the hot phase represents their ambient gas, shock-heated by SN explosions. $T_h$ evolves from $T_h > 10^6$~K down to  $T_h  \sim 10^4$~K by a molecule- and metallicity-dependent cooling function until the gas becomes neutral and flows into the cold phase. The two phases can then exchange mass and energy through three main feedback processes:  formation of stellar mass in cold clouds, cloud evaporation induced by supernova explosions and cloud growth caused by thermal instabilities through cooling of the hot phase.  

Star formation is modeled as a self-regulated, ``quiescent'' process that converts cold clouds into stars on a characteristic time-scale $t_\star$.
At each time step $dt$ a mass fraction $\beta$ of these stars are assumed to evolve by accounting for the assumed IMF  and the mass-dependent lifetime of supernova progenitors. As a result, the cold phase is depleted at a rate  $\rho_{\rm c}/t_\star$, while the hot phase increases due to gas (enriched with metals) returned by SN explosions by a term equal to $A \, \beta \, \rho_{\rm c}/t_\star$, where $A$ is an efficiency factor that
quantifies cloud evaporation inside the hot bubbles of exploding SN\footnote{The factors $A$ and the time-scale $t_\star$ are assumed to depend only on density as: $A=A_0 (\rho/\rho_{\rm th})^{-4/5}, t_{\star} = t_0^\star (\rho/\rho_{\rm th})^{-1/2}$,
where the parameters $A_0$ and $t_0^\star$ are set to $10^{3}$ and 2.1 Gyr. See \citep{2003MNRAS.339..289S} for more details.}.

Cloud formation  and growth (leading to a mass transfer from the ambient gas) are finally due to a thermal instability throughout radiative losses (see \citealt{2003MNRAS.339..289S} and references therein). Thermal instability is assumed to occur only when the density is above a given threshold value, $\rho > \rho_{\rm th}$\footnote{In section \ref{sec:galForm}  we introduced the threshold value adopted in the simulation in terms of the number density of the gas, $n_{\rm th} =  \rho_{\rm th}/(\mu m_{\rm H})$, where $\mu$ is the mean molecular weight of the gas and $m_{\rm H}$ is the mass of the hydrogen nucleus.}, and it is controlled by the parameter $f$. When $\rho > \rho_{\rm th}$, $f = 0$ and the mechanism operates creating clouds, otherwise $f = 1$. Finally note that during their growth, the volumes of the cold and hot phases are assumed constant. 

Under the previous assumptions, the evolution of the hot and cold phases throughout star formation, evolution and mass exchange are then described by the following coupled equations:
\begin{eqnarray}
\frac{d \rho_{\rm c}}{dt} & = & -\frac{\rho_{\rm c}}{t_\star} - A \beta \frac{\rho_{\rm c}}{t_\star} + \frac{1-f}{u_{\rm h} - u_{\rm c}} \Lambda (\rho_{\rm h}, u_{\rm h}), \\
\frac{d \rho_{\rm h}}{dt} & = & \beta \frac{\rho_{\rm c}}{t_\star} + A \beta \frac{\rho_{\rm c}}{t_\star} - \frac{1-f}{u_{\rm h} - u_{\rm c}} \Lambda (\rho_{\rm h}, u_{\rm h}),
\label{eq:massexchange}
\end{eqnarray}
\noindent
where $\epsilon_{\rm SN}$ is the IMF-averaged energy released by SNe per unit stellar mass formed. The evolution of the temperature of the hot component is given instead by the equation of energy conservation:
\begin{equation}
 \frac{d}{dt}(\rho_{\rm h} u_{\rm h} +\rho_{\rm c} u_{\rm c}) = \beta \frac{\rho_{\rm c}}{t_{\star}} u_{\rm SN} - \Lambda (\rho_{\rm h}, u_{\rm h}) - (1-\beta) \frac{\rho_{\rm c}}{t_{\star}} u_{\rm c},
\label{eq:energy}
\end{equation}
\noindent
where the first term describes the heating rate arising from supernovae, the second term accounts for the radiative losses of the hot phase, and the third term describes the loss of energy caused by the transformation of gas into stars, that are assumed to be at the temperature of the cold clouds. Note that in this self-regulated star formation, the value of $u_{\rm h}$ is determined only by the effects of star formation and feedback (see equations 8 - 10 in \citealt{2003MNRAS.339..289S}).

In the implementation of dust evolution, we often refer to the mass fraction of cold clouds $x_c$, defined as: $x_c \equiv \rho_{\rm c}/\rho$.

\section{Properties of models where dust evolution has been implemented}
\label{sec:APPB_ThModels}

Here we simply collect the efficiency parameters adopted by the theoretical models described in Section \ref{sec:ThModels}. Whenever the models adopt different scaling values of gas number density, temperature and metallicity, here they are aligned to the formulas in Section \ref{sec:dustGadget} to allow a direct comparison. Table \ref{tab:modelValues} collects the above data and can be used to understand differences discussed in Section \ref{sec:results}.
\end{appendix}

\label{lastpage}
\end{document}